\pdfoutput=1
\RequirePackage{ifpdf}
\ifpdf 
\documentclass[pdftex]{sigma}
\else
\documentclass{sigma}
\fi

\numberwithin{equation}{section}

\newtheorem{Theorem}{Theorem}[section]
{ \theoremstyle{definition}
\newtheorem{Remark}[Theorem]{Remark} }

\newcommand{\pa}{\partial}

\begin{document}

\newcommand{\arXivNumber}{1810.13368}

\renewcommand{\PaperNumber}{037}

\FirstPageHeading

\ShortArticleName{Generalised Darboux--Koenigs Metrics and 3-Dimensional Superintegrable Systems}

\ArticleName{Generalised Darboux--Koenigs Metrics\\ and 3-Dimensional Superintegrable Systems}

\Author{Allan P.~FORDY~$^\dag$ and Qing HUANG~$^\ddag$}

\AuthorNameForHeading{A.P.~Fordy and Q.~Huang}

\Address{$^\dag$~School of Mathematics, University of Leeds, Leeds LS2 9JT, UK}
\EmailD{\href{mailto:A.P.Fordy@leeds.ac.uk}{A.P.Fordy@leeds.ac.uk}}

\Address{$^\ddag$~School of Mathematics, Northwest University, Xi'an 710069, People's Republic of China}
\EmailD{\href{mailto:hqing@nwu.edu.cn}{hqing@nwu.edu.cn}}

\ArticleDates{Received November 01, 2018, in final form April 16, 2019; Published online May 05, 2019}

\Abstract{The Darboux--Koenigs metrics in 2D are an important class of conformally flat, non-constant curvature metrics with a single Killing vector and a pair of quadratic Killing tensors. In [arXiv:1804.06904] it was shown how to derive these by using the conformal symmetries of the 2D Euclidean metric. In this paper we consider the conformal symmetries of the 3D Euclidean metric and similarly derive a large family of conformally flat metrics possessing between 1 and 3 Killing vectors (and therefore not constant curvature), together with a~number of quadratic Killing tensors. We refer to these as {\em generalised Darboux--Koenigs metrics}. We thus construct multi-parameter families of super-integrable systems in 3 degrees of freedom. Restricting the parameters increases the isometry algebra, which enables us to fully determine the Poisson algebra of first integrals. This larger algebra of isometries is then used to reduce from 3 to 2 degrees of freedom, obtaining Darboux--Koenigs kinetic energies with potential functions, which are specific cases of the known super-integrable potentials.}


\Classification{17B6; 37J15; 37J35; 70G45; 70G65; 70H06}

\section{Introduction}
In recent years there has been a burst of activity in the identification and classification of super-integrable systems, both classical and quantum (see the review~\cite{13-2} and references therein). Most of the interest is in Hamiltonians which are in ``natural form'' (the sum of kinetic and potential energies), with the kinetic energy being {\em quadratic} in momenta. A non-degenerate kinetic energy is associated with (pseudo-)Riemannian metric, so the leading order term in any integral defines a Killing tensor for this metric, and itself commutes with the kinetic energy.

For a manifold with coordinates $(q_1,\dots ,q_n)$, metric coefficients $g_{ij}$, with inverse $g^{ij}$, the geodesic equations are Hamiltonian, with {\em kinetic energy}
\begin{gather}\label{Ham-h2}
H = \frac{1}{2} \sum_{i,j=1}^n g^{ij}p_ip_j,\qquad\mbox{where}\quad p_i=\sum_k g_{ik}\dot q_k.
\end{gather}
For a metric with isometries, the infinitesimal generators (Killing vectors) give rise to first integrals, which are {\em linear} in momenta (Noether constants).

The most common examples found in the literature have kinetic energies which are associated with flat or constant curvature metrics, so finding the leading order parts of higher order integrals (the Killing tensors) is straightforward, since, as is well known \cite{74-7}, they are all built as polynomial expressions in the {\em first order integrals} (Killing vectors). Indeed, in this case, (\ref{Ham-h2}) is actually the second order {\em Casimir} function of the symmetry algebra. However, for the {\em non-constant curvature} case, there is no algorithmic way known for building Killing tensors. In \cite{f18-4} a new method was proposed for the ``next'' class of metrics, namely the {\em conformally flat} case.

{\em Conformally flat} metrics possess an algebra of {\em conformal symmetries}, corresponding to the particular flat metric to which they are related. In 2 dimensions, as is well known, the conformal algebra is {\em infinite}. For $n\ge 3$ this algebra is {\em finite} and has {\em maximal} dimension $\frac{1}{2} (n+1)(n+2)$, which is achieved for {\em conformally flat} spaces (which includes {\em flat} and {\em constant curvature} spaces). Any two conformally equivalent metrics have the same conformal algebra, so we can describe this in terms of the corresponding {\em flat} metric. In flat spaces, the infinitesimal generators consist of $n$ {\em translations}, $\frac{1}{2} n (n-1)$ {\em rotations}, 1 {\em scaling} and $n$ {\em inversions}, totalling $\frac{1}{2} (n+1)(n+2)$. This algebra is isomorphic to $\mathfrak{so}(n+1,1)$ (see \cite[Vol.~1, p.~143]{84-4}). Whilst the conformal algebra in two-dimensional spaces is {\em infinite}, there still exists the 6-dimensional subalgebra described above (with $n=2$).

The {\em true symmetries} (isometries) of a metric form a subalgebra of the conformal algebra. The maximum dimension of the algebra of isometries is $\frac{1}{2}n(n+1)$, which is realised when the space is flat or constant curvature. To discuss {\em conformally flat} metrics, we wish to avoid the flat and constant curvature cases, so should not have ``too many isometries''. In 2 dimensions we cannot have more than {\em one} Killing vector, since, by a theorem of Darboux and Koenigs, if such a metric possesses at least {\em two} Killing vectors, then it possesses {\em three} and the space has {\em constant curvature} (see \cite{03-11,02-6}). Killing vectors correspond to first integrals which are {\em linear} in momenta. Metrics found by Koenigs (see \cite{03-11,02-6, 72-5}) are characterised by having one {\em linear} and two {\em quadratic} first integrals. In modern day jargon, these are {\em super-integrable} systems, defined on spaces which are {\em conformally flat}, but {\em not} constant curvature. In \cite{f18-4} these systems were rederived by a new approach which builds higher order {\em integrals} as polynomial expressions in {\em conformal symmetries}, together with an important {\em non-constant} multiple of the Hamiltonian itself (see Section \ref{quadrat-int} below). The Darboux--Koenigs metrics have been further generalised in~2 dimensions, by considering Hamiltonians with one {\em linear} and one or more {\em higher order} first integrals (see \cite{11-3} for the cubic case and \cite{17-4} for integrals of {\em any} integer order).

In this paper, we consider the generalisation to 3 degrees of freedom. There exist $N$-dimensional generalisations of Darboux--Koenigs metrics, both in the classical and quantum domain \cite{09-10,11-4}. In these papers the approach is more geometric and the connection to the 2-dimensional Darboux--Koenigs metrics is different from the one presented below.

In Section~\ref{conformal} we describe the structure of the 10-dimensional conformal algebra in this case, and show that it has a particular decomposition introduced in \cite{f18-1}. We show that this algebra has 4 involutive automorphisms, which play an important role in our calculations. In Section \ref{quadrat-int} we describe the method introduced in \cite{f18-4} for the construction of (in this paper) quadratic first integrals out of conformal symmetries. We initially only assume the existence of {\em one} Killing vector~$K$, but in 3 dimensions we can have {\em up to $4$ Killing vectors} without forcing constant curvature (although the examples in this paper have at most 3). The calculations give rise to metrics which depend upon a number of parameters, and degeneration of these parameters leads to metrics with additional Killing vectors (``over-degeneration'' leading to flat or constant curvature metrics). The significance of the additional Killing vectors is that we can use these to reduce from~3 to~2 degrees of freedom, using the ``Kaluza--Klein reduction'' approach described in~\cite{f18-2}. This process leads to Hamiltonians in~2 degrees of freedom with one {\em linear} and two {\em quadratic} first integrals, which are Darboux--Koenigs kinetic energies with the addition of {\em potential functions}, which are compared with those of~\cite{03-11,02-6}. With enough Killing vectors, we can reduce in different ways, leading to the {\em same} Darboux--Koenigs kinetic energy, but {\em different} potentials. Within the 3-dimensional framework we give point transformations between these.

Another consequence of having more Killing vectors is that the Poisson algebra of first integrals is simplified, so that we can find the {\em full set of relations}. The involutive automorphisms play a crucial role here, since each algebra is invariant with respect to one of these.

On the other hand, an example treated in Section~\ref{F1=e2h2+:beta=0-red} has only {\it one} Killing vector, so reduction to 2 dimensions results in a metric with {\it no} Killing vectors, but with higher order first integrals, thus leading to a maximally super-integrable system which falls out of the Darboux--Koenigs and later classifications.

The results on Poisson algebras are given in Sections \ref{sec:K=e1} to \ref{sec:K=h4}, whilst those on reductions can be found in Sections \ref{sec:F1=e2f2+red} to \ref{sec:h4-F1e12-red}.

\section{The 3D Euclidean metric and its conformal algebra}\label{conformal}

Consider metrics which are conformally related to the standard Euclidean metric in 3 dimensions, with Cartesian coordinates $(q_1,q_2,q_3)$. The corresponding kinetic energy~(\ref{Ham-h2}) takes the form
\begin{gather}\label{3d-gen}
H = \varphi(q_1,q_2,q_3) \big(p_1^2+p_2^2+p_3^2\big).
\end{gather}
A {\em conformal invariant} $X$, linear in momenta, will satisfy $\{X,H\}=\lambda(X) H$, for some function $\lambda(X)$. The conformal invariants form a Poisson algebra, which we call the {\em conformal algebra}. For special cases of $\varphi(q_1,q_2,q_3)$ there will be a subalgebra for which $\{X,H\}=0$, thus forming {\em true invariants} of $H$. These correspond to {\em infinitesimal isometries} (Killing vectors) of the metric. Constant curvature metrics possess $\frac{1}{2} n(n+1) = 6$ Killing vectors (when $n=3$).

As discussed in the introduction, the conformal algebra of this 3D metric has dimension $\frac{1}{2} (n+1)(n+2)=10$ (when $n=3$). A convenient basis is as follows
\begin{subequations}\label{g1234}
\begin{gather}
e_1 = p_1,\qquad h_1 = -2(q_1p_1+q_2p_2+q_3p_3),\nonumber\\
f_1=\big(q_2^2+q_3^2-q_1^2\big)p_1-2q_1q_2p_2-2q_1q_3p_3, \label{g1}\\
 e_2 = p_2,\qquad h_2=2(q_1p_2-q_2p_1),\nonumber\\
 f_2 =-4q_1q_2p_1-2\big(q_2^2-q_1^2-q_3^2\big)p_2-4q_2q_3p_3, \label{g2}\\
e_3 = p_3,\qquad h_3 = -2q_3p_1+2q_1p_3,\nonumber\\
 f_3=-4q_1q_3p_1-4q_2q_3p_2-2\big(q_3^2-q_1^2-q_2^2\big)p_3, \label{g3}\\
h_4 = 4q_3p_2-4q_2p_3. \label{g4}
\end{gather}
\end{subequations}
The Poisson relations of the ten elements in the conformal algebra (\ref{g1234}) are given in Table~\ref{Tab:g1234}.
\begin{table}[h]\centering
\caption{The 10-dimensional conformal algebra (\ref{g1234}).}\label{Tab:g1234}\vspace{1mm}
{\footnotesize
\renewcommand\arraystretch{1.26}\begin{tabular}{|c||c|c|c||c|c|c||c|c|c||c|}
\hline &$e_1$ &$h_1$ &$f_1$ &$e_2$ &$h_2$ &$f_2$ &$e_3$ &$h_3$ &$f_3$ &$h_4$\\[.10cm]\hline\hline
$e_1$ &0 &$2e_1$ &$-h_1$ &0 &$-2e_2$ &$-2h_2$ &0 &$-2e_3$ &$-2h_3$ &0\\[1mm]\hline
$h_1$ & &0 &$2f_1$ &$-2e_2$ &0 &$2f_2$ &$-2e_3$ &0 &$2f_3$ &0\\[1mm]\hline
$f_1$ & & &0 &$-h_2$ &$-f_2$ &0 &$-h_3$ &$-f_3$ &0 &0\\[1mm]\hline\hline
$e_2$ & & & &0 &$2e_1$ &$-2h_1$ &0 & 0 &$h_4$ &$4e_3$\\[1mm]\hline
$h_2$ & & & & &0 &$-4f_1$ & 0 &$-h_4$ &0 &$4h_3$\\[1mm]\hline
$f_2$ & & & & & &0 &$h_4$ &0 &0 &$4f_3$\\[1mm]\hline\hline
$e_3$ & & & & & & &0 &$2e_1$ &$-2h_1$ &$-4e_2$\\[1mm]\hline
$h_3$ & & & & & & & &0 &$-4f_1$ &$-4h_2$\\[1mm]\hline
$f_3$ & & & & & & & & &0 &$-4f_2$\\[1mm]\hline\hline
$h_4$ & & & & & & & & & &0\\[1mm]\hline
\end{tabular}
}\end{table}
Note that this is an example of the conformal algebra given in \cite[Table 3]{f18-1} corresponding to the case $a_2=a_3=2$, $a_4=0$. The subalgebra $\mathfrak{g}_1$, with basis (\ref{g1}) is just a copy of $\mathfrak{sl}(2)$. We then make the {\em vector space decomposition} of the full algebra $\mathfrak{g}$ into invariant subspaces under the action of $\mathfrak{g}_1$:
\begin{gather*}
\mathfrak{g} = \mathfrak{g}_1 + \mathfrak{g}_2 + \mathfrak{g}_3 + \mathfrak{g}_4.
\end{gather*}
The basis elements for $\mathfrak{g}_i$ have the same subscript and are given in the rows of~(\ref{g1234}).

The algebra (\ref{g1234}) possesses a number of involutive automorphisms:
\begin{gather*}
\iota_{12}\colon \ (q_1,q_2,q_3)\mapsto(q_2,q_1,q_3),\qquad \iota_{13}\colon \ (q_1,q_2,q_3)\mapsto(q_3,q_2,q_1),\\ \iota_{23}\colon \ (q_1,q_2,q_3)\mapsto(q_1,q_3,q_2),\\
 \iota_{ef}\colon \ (q_1,q_2,q_3)\mapsto \left(-\frac{q_1}{q_1^2+q_2^2+q_3^2},-\frac{q_2}{q_1^2+q_2^2+q_3^2},-\frac{q_3}{q_1^2+q_2^2+q_3^2}\right),
\end{gather*}
whose actions are given in Table \ref{Tab:Invg1234}.
\begin{table}[h]\centering
\caption{The involutions of the conformal algebra (\ref{g1234}).}\label{Tab:Invg1234}\vspace{1mm}
{\footnotesize\renewcommand\arraystretch{1.26}\begin{tabular}{|c||c|c|c||c|c|c||c|c|c||c|}
\hline
 &$e_1$ &$h_1$ &$f_1$ &$e_2$ &$h_2$ &$f_2$ &$e_3$ &$h_3$ &$f_3$ &$h_4$ \\[.10cm]\hline\hline
$\iota_{12}$ &$e_2$ &$h_1$ &$\frac{1}{2}f_2$ &$e_1$ &$-h_2$ &$2 f_1$ &$e_3$ &$-\frac{1}{2}h_4$ &$f_3$ &$-2h_3$ \\[1mm]\hline
$\iota_{13}$ &$e_3$ &$h_1$ &$\frac{1}{2}f_3$ &$e_2$ &$\frac{1}{2}h_4$ &$f_2$ &$e_1$ &$-h_3$ &$2f_1$ &$2h_2$ \\[1mm]\hline
$\iota_{23}$ &$e_1$ &$h_1$ &$f_1$ &$e_3$ &$h_3$ &$f_3$ &$e_2$ &$h_2$ &$f_2$ &$-h_4$ \\[1mm]\hline
$\iota_{ef}$ &$-f_1$ &$-h_1$ &$-e_1$ &$-\frac{1}{2}f_2$ &$h_2$ &$-2e_2$ &$-\frac{1}{2}f_3$ &$h_3$ &$-2e_3$ &$h_4$\\[1mm]\hline
\end{tabular}}
\end{table}

\subsection[Different choices for the decomposition of $\mathfrak{g}$]{Different choices for the decomposition of $\boldsymbol{\mathfrak{g}}$}\label{conformal-gij}

The first block of Table~\ref{Tab:Invg1234} shows different copies of $\mathfrak{sl}(2)$, which we might have chosen as our $\mathfrak{g}_1$. The remaining blocks are just the corresponding invariant subspaces.

In \cite{f18-1} we showed that the $6$-dimensional algebra $\mathfrak{g}_1 + \mathfrak{g}_2$ (which here we call $\mathfrak{g}^{(0)}$) has two quadratic Casimirs
\begin{subequations}\label{casimirs}
\begin{gather}\label{g0}
{\cal C}_2^{(0)} = 4 e_1 f_1 +h_1^2+2 e_2 f_2 - h_2^2,\qquad {\cal C}_4^{(0)} = h_1 h_2+e_1 f_2 -2 e_2 f_1,
\end{gather}
the second of which is derived from a quartic Casimir, which happens to be a perfect square in this case. In the $6\times 6$ matrix representation (the adjoint representation), this is a multiple of the identity
matrix, but in our Poisson representation, it vanishes identically, so the 6-dimensional Poisson algebra has a quadratic constraint.

Under the action of the involutions, this $6$-dimensional algebra is transformed into the first 6 elements found to the right of $\iota_{ij}$ in Table \ref{Tab:Invg1234} (here labelled $\mathfrak{g}^{(ij)}$), and the Casimirs take the form
\begin{gather}
 {\cal C}_2^{(12)} = {\cal C}_2^{(0)},\qquad {\cal C}_4^{(12)} =-{\cal C}_4^{(0)}, \label{g12} \\
 {\cal C}_2^{(13)} = 2 e_2 f_2+2 e_3 f_3+h_1^2 -\tfrac{1}{4} h_4^2,\qquad {\cal C}_4^{(13)} =e_3f_2-e_2f_3+\tfrac{1}{2} h_1 h_4, \label{g13} \\
 {\cal C}_2^{(23)} = 4 e_1 f_1 +h_1^2+2 e_3 f_3 - h_3^2,\qquad {\cal C}_4^{(23)} = h_1 h_3+e_1 f_3 -2 e_3 f_1, \label{g23} \\
 {\cal C}_2^{(ef)} = {\cal C}_2^{(0)},\qquad {\cal C}_4^{(ef)} = -{\cal C}_4^{(0)}. \label{gef}
\end{gather}
Each of the Casimirs ${\cal C}_2^{(0)}$ and ${\cal C}_2^{(ij)}$ represent spaces of constant curvature, whilst each of ${\cal C}_4^{(0)}$ and ${\cal C}_4^{(ij)}$ identically vanishes, so is a {\em quadratic constraint} on the {\em whole} $10$-dimensional conformal algebra.

There are 2 more $6$-dimensional algebras:
\begin{gather}\label{ghe}
\mathfrak{g}^{(he)}=\langle h_2,h_3,h_4,e_1,e_2,e_3\rangle, \qquad \mathfrak{g}^{(hf)}=\langle h_2,h_3,h_4,f_1,f_2,f_3\rangle ,
\end{gather}
with Casimirs
\begin{gather}
{\cal C}_2^{(he)} = e_1^2+e_2^2+e_3^2,\qquad {\cal C}_4^{(he)} = e_1 h_4+2 e_2 h_3-2 e_3 h_2, \label{cashe} \\
 {\cal C}_2^{(hf)} = f_1^2+\tfrac{1}{4} \big(f_2^2+f_3^2\big),\qquad {\cal C}_4^{(hf)} = f_3 h_2-f_2 h_3-f_1 h_4, \label{cashf}
\end{gather}
\end{subequations}
which are related through the involution $\iota_{ef}$. Again, ${\cal C}_4^{(he)}$ and ${\cal C}_4^{(hf)}$ vanish identically.

\section{Geodesic flows in 3D with linear and quadratic integrals}\label{quadrat-int}

In this and later sections we seek Hamiltonian functions (\ref{3d-gen}) which possess up to 4 first order invariants (Killing vectors), together with a number of quadratic integrals. In 3 dimensions, if we have 5 isometries, then we {\it must} have 6, so the space has constant curvature. We already mentioned that in 2 dimensions, Darboux and Koenigs proved that we can only have 1 isometry (i.e., 2 implied~3). This ``gap phenomenon'' occurs in all dimensions and is the subject of~\cite{14-7}.

Initially we only assume the existence of {\em one} Killing vector $K$, the calculations giving rise to metrics which depend upon a number of parameters. Degeneration of these parameters leads to metrics with additional Killing vectors, with ``over-degeneration'' leading to flat or constant curvature metrics. We use the method introduced in \cite{f18-4} to construct {\em quadratic invariants} out of conformal invariants.

A {\em quadratic conformal invariant} is any expression of the form
\begin{gather}\label{gen-quad}
F = \sum_{i,j=1}^{10} \beta_{ij} X_i X_j + \psi(q_1,q_2,q_3) H,
\end{gather}
where $\beta_{ij}$ is a symmetric matrix of ({\em constant}) coefficients, $X_i$ are {\em linear} conformal invariants, and $\psi(q_1,q_2,q_3)$ is an arbitrary function, which satisfies
\begin{gather*}
\{F,H\} = \left(\sum_{i=1}^3 \mu_i(q_1,q_2,q_3) p_i\right) H,
\end{gather*}
where $\mu_i(q_1,q_2,q_3)$ are some functions.

We can ask whether there is a choice of $\beta_{ij}$ and $\psi(q_1,q_2,q_3)$ for which $\mu_i(q_1,q_2,q_3)\equiv 0$, in which case {\em $F$ is a quadratic invariant}. In fact, we have more structure. If {\em both} $K$ {\em and} $F$ are invariants, then so are $\{F,K\}$, $\{\{F,K\},K\}$, etc.

\subsection{Using the involutions}

It follows from the involutions of Table \ref{Tab:Invg1234}, that we can limit our choice of a {\em single first order invariant} to just 3 elements of our algebra: $e_1, h_1$ and $h_4$. Any other element is equivalent to one of these. Having fixed this invariant we choose a {\em quadratic conformal invariant} (\ref{gen-quad}) with specific form
\begin{gather*}
F = X_i X_j + \psi(q_1,q_2,q_3) H,
\end{gather*}
where $X_i$ and $X_j$ are {\em linear} conformal invariants. At this stage we still have some equivalent choices from involutions which leave our choice of Killing vector invariant. For example, $e_1$ is fixed under the action of $\iota_{23}$, while $h_1$ is fixed (up to sign) by them all and $h_4$ is fixed (up to sign) by $\iota_{23}$ and $\iota_{ef}$.

\subsection{Reductions to two dimensions}\label{sec:reductions}

In Sections \ref{sec:K=e1}, \ref{sec:K=h1} and \ref{sec:K=h4} we present a number of systems in 3 degrees of freedom, which have {\em at least one} first integral of {\em degree one} in momenta, as well as 3 independent {\em quadratic} integrals. These systems depend upon a number of parameters and, in full generality, the Poisson algebras are complicated, but by restricting parameters we can obtain systems with larger isometry algebras (first order integrals) and this enables us to calculate the full set of Poisson relations. The involutions play an important role in this calculation.

There is clearly an analogy with the Darboux--Koenigs systems in 2 degrees of freedom, but we show that there is a much deeper connection. The increased isometry algebra, obtained by restricting parameters, can be used to {\em reduce} the system from 3 to 2 degrees of freedom. After this process, we find that each of our systems reduces to one particular Darboux--Koenigs system, but with the addition of a potential. We compare the resulting potentials with those presented in \cite{03-11,02-6}.

We use the particular method of reduction introduced in \cite{f18-2}, referred to as the ``Kaluza--Klein reduction'', since it is essentially the reverse procedure to the Kaluza--Klein {\it extension}. By adapting coordinates to a linear first integral, we can reduce from 3 to 2 degrees of freedom. In principle, the lower-dimensional system would possess {\em vector potential} terms, but in the examples of this paper (and of \cite{f18-2}) these can be removed by gauge transformation. Reduction to 2 degrees of freedom is not enough. We would like our system to be at least {\em completely integrable} and, preferably, {\em maximally super-integrable}. This requires the existence of integrals of the original system, which {\em commute} with the linear integral. We would like to choose our coordinate system so that the reduced {\em kinetic energy} is explicitly conformally related to the standard Cartesian kinetic energy. We would like to be able to construct any {\em symmetries} and a 6-dimensional subalgebra of {\em conformal symmetries} of the corresponding $2$-dimensional metric from invariant combinations of the conformal algebra (\ref{g1234}). All of these things will be done for the systems discussed in this paper.

Each additional Killing vector allows us to reduce to 2 dimensions in a different way. From a given starting point, we always arrive at the same Darboux--Koenigs metric, but the specific reduction gives us a particular potential function. This allows us to relate {\em different} 2-dimensional potentials by a~``rotation'' in 3~space.

\section[Systems with $K=e_1$]{Systems with $\boldsymbol{K=e_1}$}\label{sec:K=e1}

In the case $K=e_1$, we consider the Hamiltonian (\ref{3d-gen}) of the form
\begin{gather}\label{H3d-e1}
H = \varphi(q_2,q_3) \big(p_1^2+p_2^2+p_3^2\big),
\end{gather}
and exploit the following chains of length 3, found in the algebra:
\begin{gather*}
\{ \cdot , e_1\} \colon \ f_1 \mapsto h_1 \mapsto -2 e_1 \mapsto 0 \qquad\mbox{and}\qquad \{ \cdot , e_1\} \colon \ f_k \mapsto 2 h_k \mapsto 4 e_k \mapsto 0,\qquad k=2,3.
\end{gather*}
We use this to build 3 functions which satisfy $\{ \cdot , e_1\} \colon F_1 \mapsto F_2 \mapsto F_3 \mapsto 0$. If we can find $\varphi$ and $F_1$, such that $\{F_1,H\}=0$, then, by the Jacobi identity, we automatically have $\{F_2,H\}=\{F_3,H\}=0$.

\subsection[$F_1 = e_2f_2+\psi(q_1,q_2,q_3) H$]{$\boldsymbol{F_1 = e_2f_2+\psi(q_1,q_2,q_3) H}$}\label{sec:F1=e2f2+}

We consider a metric (\ref{H3d-e1}) with the chain of functions generated by $F_1 = e_2f_2+\psi(q_1,q_2,q_3) H$, under the action of $e_1$:
$F_2=\frac 12 \{F_1,e_1\}$, $F_3=\frac 12 \{F_2,e_1\}$, namely,
\begin{gather}\label{F123}
F_1=e_2f_2+\psi H,\qquad F_2 = e_2 h_2+\tfrac{1}{2} \psi_{q_1} H,\qquad F_3=e_2^2+\tfrac{1}{4}\psi_{q_1q_1}H.
\end{gather}
The conditions $\{F_1,H\}=\{F_3,e_1\}=0$ lead to
\begin{gather}\label{phipsi}
 \varphi=\frac{q_2^2q_3^2}{\alpha q_2^2+\beta q_3^2+\gamma q_2^2q_3^2},\qquad \psi=-2\beta \frac{q_1^2+q_3^2}{q_2^2}+2\gamma q_2^2.
\end{gather}
Since $H$, $e_1$, $F_1$, $F_2$, $F_3$ are independent first integrals of $H$, this Hamiltonian is maximally {\em super-integrable}. We also have that $H$, $e_1$, $F_3$ are in involution.

\begin{Remark}[the St\"ackel transform]
The Hamiltonian (\ref{H3d-e1}), with $\varphi$ given by (\ref{phipsi}) can be regarded as the St\"ackel transform of the Euclidean kinetic energy with potential $V=\varphi^{-1}$. Such potentials were classified in \cite{17-3} and this example corresponds to a reduction of $V_{[3,1,1]}$, given in \cite[Section~5]{17-3}, by equating $(q_1,q_2,q_3)=(x,y,z)$ and $(\alpha,\beta,\gamma)=(a_3,a_2,a_5)$.
\end{Remark}

Notice that the resulting metric is invariant under the involution $\iota_{23}$ (extended to include $\alpha \leftrightarrow \beta$), so we also have the integrals $G_i = \iota_{23}(F_i)$. These are not, of course, functionally independent, but could be useful when considering the Poisson algebra with the involution $\iota_{23}$. In fact, $G_3$ satisfies the simple relation
\begin{gather*}
G_3+F_3= \gamma H-e_1^2.
\end{gather*}
The Poisson algebra for the general case seems to be quite complicated. Restricting the parameters introduces additional isometries. Setting {\em any two} of the parameters to zero reduces the metric to a flat or constant curvature case. However,
\begin{enumerate}\itemsep=0pt
 \item Setting just $\alpha = 0$, we have the isometry algebra $\langle e_1,e_3,h_3\rangle $, satisfying Euclidean relations.
 \item Setting just $\beta = 0$, we have the isometry algebra $\langle e_1,e_2,h_2\rangle $, satisfying Euclidean relations. This case is related to that of $\alpha=0$ by the action of the involution $\iota_{23}$.
 \item Setting just $\gamma = 0$, we have the isometry algebra $\langle e_1,h_1,f_1\rangle $, satisfying the relations of~$\mathfrak{sl}(2)$.
\end{enumerate}

\subsubsection[The Poisson algebra of integrals when $\alpha=0$]{The Poisson algebra of integrals when $\boldsymbol{\alpha=0}$}\label{sec:F1=e2f2+:a=0:pb}

Now we have
\begin{subequations}
\begin{gather}\label{F1=e2f2+:alpha=0:phipsi}
\varphi=\frac{q_2^2}{\beta + \gamma q_2^2} \qquad\mbox{and}\qquad \psi=-2\beta \frac{q_1^2+q_3^2}{q_2^2}+2\gamma q_2^2,
\end{gather}
so we have lost the involutive symmetry (so no integrals $G_i$), but gained additional Killing vectors, with isometry algebra $\langle e_1,e_3,h_3\rangle $, satisfying the Euclidean relations
\begin{gather}\label{F1=e2f2+:alpha=0:e13h3}
\{e_1,e_3\}=0,\qquad \{e_1,h_3\}= -2 e_3,\qquad \{e_3,h_3\}= 2 e_1.
\end{gather}
\end{subequations}
If we define $K_1=e_1$, $K_2=e_3$, $K_3=h_3$, then $\{K_i,F_j\}$ are easily determined, requiring only the introduction of one additional element $F_4 = e_2h_4-\frac{4 \beta q_3}{q_2^2} H$, when calculating $\{K_2,F_1\}$ and $\{K_3,F_2\}$. The formula for $F_4$ is easily determined from Table~\ref{Tab:g1234}, by considering $\{e_3,e_2 f_2\}=-e_2 h_4$. The cubic expressions $\{F_i,F_j\}$ can all be written in terms of our linear and quadratic functions. The full set of Poisson relations is given in Table~\ref{Tab:alpha=0:Pij}. Adding $H$, we have an 8-dimensional algebra obeying constraints:
\begin{gather}
I_1=K_1^2+K_2^2+F_3-\gamma H=0,\qquad I_2=K_1F_4-2K_2F_2+2K_3F_3=0,\nonumber\\
I_3=4 F_2^2-8 F_1 F_3 +F_4^2-16 \beta \gamma H^2=0.\label{F1=e2f2+:alpha=0:I123}
\end{gather}
\begin{Remark}Notice that the first two expressions are related to ${\cal C}_2^{(he)}$ and ${\cal C}_4^{(he)}$ respectively (see~(\ref{cashe})). We could, in fact, use $I_1=0$ to eliminate $F_3$ and just write a $6\times 6$ table, but $F_3$ was an integral part of the definition of this case.
\end{Remark}

\begin{table}[t]\centering \caption{The Poisson algebra of first integrals when $\alpha=0$.}\label{Tab:alpha=0:Pij}\vspace{1mm}

\resizebox{\textwidth}{!}{
\renewcommand\arraystretch{1.26}\begin{tabular}{|@{\,\,}c@{\,\,}||@{\,\,}c@{\,\,}|@{\,\,}c@{\,\,}|@{\,\,}c@{\,\,}||@{\,\,}c@{\,\,}|@{\,\,}c@{\,\,}|@{\,\,}c@{\,\,}|@{\,\,}c@{\,\,}|}
\hline
 &$K_1$ &$K_2$ &$K_3$ &$F_1$ &$F_2$ &$F_3$ &$F_4$ \\[.10cm]\hline\hline
$K_1$ &$0$ &$0$ &$-2K_2$ &$-2F_2$ &$-2F_3$ &$0$ &$0$\\[0.2mm]\hline
$K_2$ & &$0$ &$2K_1$ &$-F_4$ &$0$ &$0$ &$-4F_3$\\[0.2mm]\hline
$K_3$ & & &$0$ &$0$ &$F_4$ &$0$ &$-4F_2$\\[0.2mm]\hline\hline
$F_1$ & & & &$0$ &$-2(2K_1F_1+K_3F_4)$ &$-2(2K_1F_2+K_2F_4)$ &$8(K_3F_2-K_2F_1)$\\[0.2mm]\hline
$F_2$ & & & & &$0$ &$-4K_1F_3$ &$2(2K_2F_2+2K_3F_3-K_1F_4)$\\[0.2mm]\hline
$F_3$ & & & & & &$0$ &$8K_2F_3$\\[0.2mm]\hline
$F_4$ & & & & & & &$0$ \\[0.2mm]\hline
\end{tabular}}
\end{table}

Since $H$ is invariant under the action of $\iota_{13}$, the Poisson algebra has this symmetry. The action of $\iota_{13}$ is summarised in Table~\ref{Tab:alpha=0:i13}.
\begin{table}[t]\centering
\caption{The action of $\iota_{13}$ on $H$ and its integrals.}\label{Tab:alpha=0:i13}\vspace{1mm}
{\footnotesize\renewcommand\arraystretch{1.26}\begin{tabular}{|c||c||c|c|c||c|c|c|c|}
\hline
 & $H$ &$K_1$ &$K_2$ &$K_3$ &$F_1$ &$F_2$ &$F_3$ &$F_4$ \\[.10cm]\hline\hline
$\iota_{13}$ & $H$ &$K_2$ &$K_1$ &$-K_3$ &$F_1$ &$\frac{1}{2} F_4$ &$F_3$ &$2 F_2$ \\[1mm]\hline
\end{tabular}}
\end{table}
Several of the entries in Table \ref{Tab:alpha=0:Pij} are related through this involution. We also have that the constraints (\ref{F1=e2f2+:alpha=0:I123}) transform as
\begin{gather*}
\iota_{13}\colon \ (I_1,I_2,I_3) \mapsto (I_1,-I_2,I_3).
\end{gather*}

\subsubsection[The Poisson algebra of integrals when $\gamma=0$]{The Poisson algebra of integrals when $\boldsymbol{\gamma=0}$}\label{sec:F1=e2f2+:g=0:pb}

When $\gamma=0$, the Hamiltonian $H$ is invariant under the action of $\iota_{ef}$ (as is the $3$-dimensional isometry algebra). We define $K_1=e_1$, $K_2=h_1$, $K_3=f_1$, with $F_1$, $F_2$, $F_3$ given by~(\ref{F123}). The action of $\iota_{ef}$ on $F_2$ and $F_3$ gives us two more quadratic elements, which close the algebra:
\begin{gather}\label{F45}
F_4 = f_2 h_2 -\frac{4\beta q_1 \big(q_1^2+q_2^2+q_3^2\big)}{q_2^2} H,\qquad F_5 = f_2^2 -\frac{4\beta \big(q_1^2+q_2^2+q_3^2\big)^2}{q_2^2} H.
\end{gather}
The action of $\iota_{ef}$ is summarised in Table \ref{Tab:gam=0:ief}.
\begin{table}[t]\centering
\caption{The action of $\iota_{ef}$ on $H$ and its integrals.}\label{Tab:gam=0:ief}\vspace{1mm}
{\footnotesize\renewcommand\arraystretch{1.16}\begin{tabular}{|c||c||c|c|c||c|c|c||c|c|c||c|}
\hline
 & $H$ &$K_1$ &$K_2$ &$K_3$ &$F_1$ &$F_2$ &$F_3$ &$F_4$ &$F_5$ \\[.10cm]\hline\hline
$\iota_{ef}$ & $H$ &$-K_3$ &$-K_2$ &$-K_1$ &$F_1$ &$-\frac{1}{2} F_4$ &$\frac{1}{4} F_5$ &$-2 F_2$ &$4 F_3$ \\[1mm]\hline
\end{tabular}}
\end{table}
A consequence of this is that we can {\em deduce} the entries for $\{K_i,F_4\}$ and $\{K_i,F_5\}$ (in Table~\ref{Tab:gam=0}) from those of $\{K_j,F_2\}$ and $\{K_j,F_3\}$. The entries $\{F_3,F_4\}$ and $\{F_2,F_5\}$ are similarly related. Some of the longer entries are labelled $P_{ij}$ and listed below.
\begin{table}[h]\centering
\caption{The Poisson algebra of first integrals when $\gamma=0$.}\label{Tab:gam=0}\vspace{1mm}
{\footnotesize\renewcommand\arraystretch{1.16}
\resizebox{\textwidth}{!}{\begin{tabular}{|@{\,\,}c@{\,\,}||@{\,\,}c@{\,\,}|@{\,\,}c@{\,\,}|@{\,\,}c@{\,\,}||@{\,\,}c@{\,\,}|@{\,\,}c@{\,\,}|@{\,\,}c@{\,\,}||@{\,\,}c@{\,\,}|@{\,\,}c@{\,\,}|}
\hline &$K_1$ &$K_2$ &$K_3$ &$F_1$ &$F_2$ &$F_3$ &$F_4$ &$F_5$ \\[.10cm]\hline\hline
$K_1$ &0 &$2K_1$ &$-K_2$ &$-2 F_2$ &$-2F_3$ &$0$ &$-6F_1-8 {\cal C}_K +4 (2\alpha+\beta)H$ &$-4F_4$ \\[1mm]\hline
$K_2$ & &0 &$2K_3$ &$0$ &$-2 F_2$ &$-4 F_3$ &$2 F_4$ &$4 F_5$ \\[1mm]\hline
$K_3$ & & &0 &$-F_4$ &$-3F_1-4 {\cal C}_K +2 (2\alpha+\beta)H$ &$-2 F_2$ &$-F_5$ &0\\[1mm]\hline\hline
$F_1$ & & & &0 &$8 (K_3 F_3+\beta K_1 H)$ &$4 K_2 F_3$ &$4 (K_1F_5+4 \beta K_3 H)$ & $-4 K_2 F_5$\\[1mm]\hline
$F_2$ & & & & &0 &$-4 K_1 F_3$ & $P_{24}$ &$P_{25}$ \\[1mm]\hline
$F_3$ & & & & & &0 &$P_{34}$ &$P_{35}$\\[1mm]\hline\hline
$F_4$ & & & & & & &0 &$-8K_3 F_5$ \\[1mm]\hline
$F_5$ & & & & & & & &0 \\[1mm]\hline
\end{tabular}}
}\end{table}

The longer entries are given by
\begin{gather*}
 P_{24} = 8 K_2 (-F_1-2 {\cal C}_K +(2\alpha+\beta) H),\qquad P_{25} = -4 (K_2F_4+2 K_3F_1+4\beta K_3 H), \\
 P_{34} = -4 (K_2F_2-K_1F_1-2\beta K_1 H),\qquad P_{35} = -8 K_2 (F_1+2\beta H),
\end{gather*}
where ${\cal C}_K = K_1K_3+\frac{1}{4} K_2^2$ is the Casimir function of the isometry algebra.

Adding $H$, we have a 9-dimensional algebra obeying the constraints:
\begin{gather*}
 I_1= K_1F_1+K_2F_2-2 K_3 F_3 - 2\beta K_1 H = 0 ,\\
 I_2 = K_1F_5+K_2F_4-2 K_3 F_1 + 4\beta K_3 H = 0 , \\
 I_3=F_1 F_2-F_3 F_4+2 K_2 (K_1 F_1+K_2 F_2-2 K_3 F_3) -2 \beta H F_2 = 0, \\
 I_4=F_1 F_4-F_2 F_5+2 K_2 (K_1 F_5+K_2 F_4-2 K_3 F_1) -2 \beta H F_4 = 0.
\end{gather*}
\begin{Remark}Notice that the first two expressions are related to the Casimir ${\cal C}_4^{(0)}$, of the algebra~$\mathfrak{g}^{(0)}$ of Section~\ref{conformal-gij}. Under the action of $\iota_{ef}$, they obey
\begin{gather*}
(I_1,I_2,I_3,I_4) \mapsto \big(\tfrac{1}{2} I_2,2 I_1,-\tfrac{1}{2} I_4,-2 I_3\big).
\end{gather*}
\end{Remark}

\subsection[$F_1 = e_1h_2+\psi_1(q_1,q_2,q_3) H$, $G_1 = e_1h_3+\psi_2(q_1,q_2,q_3) H$]{$\boldsymbol{F_1 = e_1h_2+\psi_1(q_1,q_2,q_3) H}$, $\boldsymbol{G_1 = e_1h_3+\psi_2(q_1,q_2,q_3) H}$}\label{sec:FGe1h2}

We again take the Hamiltonian (\ref{H3d-e1}), with first integral $K=e_1$, and consider the chains of functions generated by
\begin{subequations}\label{FGe1h2}
\begin{alignat}{3}
& F_1 = e_1h_2+\psi_1(q_1,q_2,q_3) H, \qquad && F_2 = \tfrac{1}{2} \{F_1,e_1\} = e_1e_2 + \tfrac{1}{2} \psi_{1\,q_1} H, & \label{Fe1h2}\\
& G_1 = e_1h_3+\psi_2(q_1,q_2,q_3) H, \qquad && G_2 = \tfrac{1}{2} \{G_1,e_1\} = e_1e_3 + \tfrac{1}{2} \psi_{2\,q_1} H,& \label{Ge1h3}
\end{alignat}
related through the involution $\iota_{23}$. The conditions $\{F_1,H\}=0$ and $\{G_1,H\}=0$ give the functions
\begin{gather}\label{phipsiFGe1h2}
 \varphi=\frac{1}{\alpha q_2+\beta q_3+\gamma},\qquad \psi_1=-\frac{\alpha}{2} q_1^2,\qquad \psi_2=-\frac{\beta}{2} q_1^2,
\end{gather}
\end{subequations}
so $\{F_2,e_1\}=-\frac{\alpha}{2} H$ and $\{G_2,e_1\}=-\frac{\beta}{2} H$. The integrals $H$, $F_1$, $F_2$, $G_1$, $G_2$ are functionally independent.

\begin{Remark}[the St\"ackel transform] Again, the Hamiltonian (\ref{H3d-e1}), but with $\varphi$ given by (\ref{phipsiFGe1h2}) can be regarded as the St\"ackel transform of the Euclidean kinetic energy with potential $V=\varphi^{-1}$. In the classification of~\cite{17-3}, this example corresponds to a reduction of $V_{[0]}$, given in \cite[Section~5]{17-3}, by equating $(q_1,q_2,q_3)=(z,x,y)$ and $(\alpha,\beta,\gamma)=(a_1,a_2,a_5)$.
\end{Remark}

This solution has a $3$-dimensional isometry algebra
\begin{subequations}\label{FGe1h2Ki}
\begin{gather}\label{KiFGe1h2}
 K_1 = e_1,\qquad K_2 = \beta e_2-\alpha e_3,\qquad K_3 = \beta h_2-\alpha h_3 ,
\end{gather}
satisfying
\begin{gather} \label{KijFGe1h2}
 \{K_1,K_2\} = 0,\qquad \{K_1,K_3\} = -2 K_2,\qquad \{K_2,K_3\} = 2 \big(\alpha^2+\beta^2\big) K_1,
\end{gather}
with Casimir
\begin{gather}\label{casKFGe1h2}
{\cal C}_K = \big(\alpha^2+\beta^2\big) K_1^2+K_2^2.
\end{gather}
\end{subequations}
We then find that
\begin{gather*}
\alpha G_1-\beta F_1+K_1 K_3= 0, \qquad\mbox{and}\qquad \alpha G_2-\beta F_2+K_1 K_2= 0,
\end{gather*}
with the integrals $H$, $K_1$, $K_2$, $K_3$, $F_1$ being functionally independent.

\begin{table}[t]\centering
\caption{The Poisson algebra of first integrals of (\ref{H3d-e1}), with (\ref{phipsiFGe1h2}).}\label{FGe1h2:PBtable}\vspace{1mm}
\addtolength{\tabcolsep}{-4.6pt}

\resizebox{\textwidth}{!}{
\renewcommand\arraystretch{1.26}\begin{tabular}{|c||c|c|c||c|c|c|c|c|c|}
\hline
 &$K_1$ &$K_2$ &$K_3$ &$F_1$ &$F_2$ &$F_3$ &$F_4$ &$F_5$ &$F_6$ \\[1.5mm]\hline\hline
$K_1$ &$0$ &$0$ &$-2K_2$ &$-2F_2$ &$\frac12\alpha H$ &$-2F_4$ &$0$ &$0$ &$0$ \\[1.5mm]\hline
$K_2$ & &$0$ &$2\delta K_1$ &$2\beta K_1^2$ &$0$ &$-4\delta F_2+8\beta K_1K_2$ &$\alpha\delta H$ &$-\alpha\beta H$ &$\bar{\alpha}(\delta F_4-2\beta K_2^2)$ \\[1.5mm]\hline
$K_3$ & & &$0$ &$F_3$ &$F_4\!-\!2\beta K_1^2$ &$4\beta K_1K_3-8\delta F_1-4\alpha F_6$ &$-4\delta F_2-4\beta K_1K_2$ &$4(\beta F_2\!+\!K_1K_2)$ &$\bar{\alpha}(\delta F_3-2\beta K_2K_3)$\\[1.5mm]\hline\hline
$F_1$ & & & &$0$ &$-2K_1F_5$ &$4K_3F_5-8\beta K_1F_1$ &$-4K_1(K_1K_2\!+\!\beta F_2)$ &$8K_1F_2$ &$\bar{\alpha} K_1(\beta F_3-2K_2K_3)$
\\[1.5mm]\hline
$F_2$ & & & & &$0$ &$4(K_1^2K_2\!-\!\beta K_1F_2\!+\!K_2F_5)$ &$\alpha\beta K_1 H$ &$-2\alpha K_1 H$ &$\bar{\alpha} K_1(\beta F_4-2K_2^2)$ \\[1.5mm]\hline
$F_3$ & & & & & &$0$ &$P_{34}$ &$P_{35}$ &$P_{36}$ \\[1.5mm]\hline
$F_4$ & & & & & & &$0$ &$-2\alpha K_2 H$ &$4\bar{\alpha} K_2(\delta (K_1^2-F_5)-K_2^2)$ \\[1.5mm]\hline
$F_5$ & & & & & & & &$0$ &$2\bar{\alpha} K_2(F_4\!+\!2\beta F_5\!-\!2\beta K_1^2)$ \\[1.5mm]\hline
$F_6$& & & & & & & & &$0$ \\[1.5mm]\hline
\end{tabular}
}
\end{table}

We can use $\langle K_1, K_2, K_3, F_1,F_2\rangle $ to generate a Poisson algebra. The action of $K_i$ on $F_i$ and the Poisson bracket $\{F_1,F_2\}$ require the introduction of four more quadratic elements:
\begin{gather*}
 F_3 = 2 K_2 h_2-\alpha K_1 h_4 -2 \alpha q_1 (\beta q_2-\alpha q_3) H,\qquad F_4 = 2 e_2 K_2 -\alpha (\beta q_2-\alpha q_3) H, \nonumber\\
 F_5=e_1^2-e_2^2+\alpha q_2H, \qquad F_6=K_2h_4+(\beta q_2-\alpha q_3)^2H, 
\end{gather*}
satisfying the relations given in Table \ref{FGe1h2:PBtable}.
In this table we have $\delta=\alpha^2+\beta^2$, $\bar{\alpha}=\frac2\alpha$ and
\begin{gather*}
P_{34} = 4 K_1\big(\beta F_4-2\alpha^2K_1^2-6K_2^2+2\alpha^2\gamma H\big),\\
P_{35} = 8 K_1\big(3F_4+\beta F_5-\beta K_1^2\big)-24K_2F_2-8\alpha K_3H,\\
P_{36} = \frac4\alpha K_3\big(\beta F_4-2\alpha^2 K_1^2-4K_2^2+2\alpha^2\gamma H\big)+\frac{4\beta}\alpha K_2F_3.
\end{gather*}
This 10-dimensional algebra (including $H$ itself) is constrained by the following relations
\begin{gather*}
 I_1 = K_1F_3+2K_3F_2-4K_2F_1=0,\qquad I_2 = 2\alpha K_1F_6+K_2F_3-K_3F_4=0, \\
I_3 = \alpha K_3H+4K_2F_2-2K_1F_4=0,\qquad I_4 = 8K_1K_2F_5-8K_1^3K_2+4F_2F_4+\alpha F_3H=0, \\
I_5 = \big(2\alpha^2+\beta^2\big)K_1^2+K_2^2-\big(\alpha^2+\beta^2\big)F_5-\beta F_4-\alpha^2\gamma H=0.
\end{gather*}

\section[Systems with $K=h_1$]{Systems with $\boldsymbol{K=h_1}$}\label{sec:K=h1}

In the case $K=h_1$, we consider the Hamiltonian (\ref{3d-gen}) of the form
\begin{gather}\label{H3d-h1}
H = \varphi(q_1,q_2,q_3)\big(p_1^2+p_2^2+p_3^2\big) = q_1^2 \Phi\left(\frac{q_2}{q_1},\frac{q_3}{q_1}\right) \big(p_1^2+p_2^2+p_3^2\big),
\end{gather}
which is the most general form commuting with $h_1$. This is also generally invariant with respect to the involution $\iota_{ef}$, so we choose our functions $F_i$ to have this invariance, noting that $\iota_{ef}\colon (e_2,h_2,f_2)\mapsto\big({-}\frac{1}{2} f_2,h_2,-2 e_2\big)$.

\subsection[$F_1 = e_2h_2+\psi_1(q_1,q_2,q_3) H$, $F_2 = f_2 h_2+\psi_2(q_1,q_2,q_3) H$, $F_3 = h_2^2+\psi_3(q_1,q_2,q_3) H$]{$\boldsymbol{F_1 = e_2h_2+\psi_1(q_1,q_2,q_3) H}$, $\boldsymbol{F_2 = f_2 h_2+\psi_2(q_1,q_2,q_3) H}$,\\ $\boldsymbol{F_3 = h_2^2+\psi_3(q_1,q_2,q_3) H}$}\label{sec:F1e2h2}

With the Hamiltonian (\ref{H3d-h1}), with first integral $K=h_1$, we consider functions of the form
\begin{gather*}
F_1 = e_2h_2+\psi_1(q_1,q_2,q_3) H, \qquad F_2 = f_2h_2 + \psi_2(q_1,q_2,q_3) H,\nonumber\\
 F_3 = h_2^2 + \psi_3(q_1,q_2,q_3) H, 
\end{gather*}
and require the conditions $\{F_i,H\}=0$ to find
\begin{gather}
 \varphi=\frac{q_2^2q_3^2\sqrt{q_1^2+q_2^2}}{\big(\alpha q_2^2+\beta q_3^2\big)\sqrt{q_1^2+q_2^2}+\gamma q_1q_3^2},\qquad
 \psi_1=-\frac{\gamma \big(2 q_1^2+ q_2^2\big)+2\beta q_1\sqrt{q_1^2+q_2^2}}{q_2^2\sqrt{q_1^2+q_2^2}}, \label{phi-e2h2}\\
 \psi_2=2\big(q_1^2+q_2^2+q_3^2\big)\psi_1,\qquad \psi_3=-\frac{4q_1\big(\beta q_1+\gamma\sqrt{q_1^2+q_2^2}\big)}{q_2^2}.\nonumber 
\end{gather}
If we extend $\iota_{ef}$ to act on the parameters: $(\alpha,\beta,\gamma)\mapsto (\alpha,\beta,-\gamma)$ then
\begin{gather*}
(H,F_1,F_2,F_3)\mapsto \big(H,-\tfrac{1}{2} F_2,-2 F_1,F_3\big).
\end{gather*}

\begin{Remark}[the St\"ackel transform] Again, the Hamiltonian (\ref{H3d-h1}), with $\varphi$ given by~(\ref{phi-e2h2}) can be regarded as the St\"ackel transform of the Euclidean kinetic energy with potential $V=\varphi^{-1}$. In the classification of~\cite{17-3}, this example corresponds to a reduction of system~iii, given in \cite[Section~7]{17-3}, by equating $(q_1,q_2,q_3)=(z,y,x)$ and $(\alpha,\beta,\gamma)=(a_3,a_4,a_5)$.
\end{Remark}

The integrals $H$, $K$, $F_1$, $F_2$, $F_3$ are functionally independent. The simpler Poisson relations are
 \begin{gather}
 \{F_1,K\}=2 F_1,\qquad \{F_2,K\}= -2 F_2,\qquad \{F_3,K\}= 0,\nonumber\\
 \{F_1,F_2\}=4 K (2 \beta H-F_3).\label{PBFK}
 \end{gather}

Again, we can restrict the parameters to increase the algebra of isometries:
\begin{enumerate}\itemsep=0pt
 \item By setting $\alpha = 0$, we have the isometry algebra $\langle e_3,h_1,f_3\rangle $, satisfying the relations of $\mathfrak{sl}(2)$.
 \item By setting $\beta = 0$, we just have the single isometry $h_1$.
 \item By setting $\gamma = 0$, the metric reduces to that of Section~\ref{sec:F1=e2f2+} (with $\gamma=0$) so has the isometry algebra $\langle e_1,h_1,f_1\rangle $.
\end{enumerate}

\subsubsection[The Poisson algebra of integrals when $\alpha=0$]{The Poisson algebra of integrals when $\boldsymbol{\alpha=0}$}\label{sec:F1=e2h2+:a=0:pb}

Now we have
\begin{gather*}
\varphi=\frac{q_2^2\sqrt{q_1^2+q_2^2}}{\beta \sqrt{q_1^2+q_2^2}+\gamma q_1},
\end{gather*}
with $\psi_i$ as before, and gain additional Killing vectors, with isometry algebra $\langle h_1,e_3,f_3\rangle $.

\begin{table}[t]\centering
\caption{The Poisson algebra of first integrals when $\alpha=0$.}\label{Tab:h1-alpha=0:Pij}\vspace{1mm}
\renewcommand\arraystretch{1.26}\begin{tabular}{|@{\,\,}c@{\,\,}||@{\,\,}c@{\,\,}|@{\,\,}c@{\,\,}|@{\,\,}c@{\,\,}||@{\,\,}c@{\,\,}|@{\,\,}c@{\,\,}|@{\,\,}c@{\,\,}|@{\,\,}c@{\,\,}|}
\hline
 &$K_1$ &$K_2$ &$K_3$ &$F_1$ &$F_2$ &$F_3$ &$F_4$ \\[.10cm]\hline\hline
$K_1$ &$0$ &$-2K_2$ &$2K_3$ &$-2F_1$ &$2F_2$ &$0$ &$0$\\[0.2mm]\hline
$K_2$ & &$0$ &$-2K_1$ &$0$ &$F_4$ &$0$ &$4F_1$\\[0.2mm]\hline
$K_3$ & & &$0$ &$F_4$ &$0$ &$0$ &$4F_2$\\[0.2mm]\hline\hline
$F_1$ & & & &$0$ &$4K_1(2\beta H-F_3)$ &$2(K_2F_4-2K_1F_1)$ &$8K_2(2\beta H-F_3)$\\[0.2mm]\hline
$F_2$ & & & & &$0$ &$2(K_3F_4+2K_1F_2)$ &$8K_3(2\beta H-F_3)$\\[0.2mm]\hline
$F_3$ & & & & & &$0$ &$-8(K_2F_2+K_3F_1)$\\[0.2mm]\hline
$F_4$ & & & & & & &$0$ \\[0.2mm]\hline
\end{tabular}
\end{table}

If we define $K_1=h_1$, $K_2=e_3$, $K_3=f_3$, then $\{K_i,F_j\}$ are easily determined, requiring only the introduction of one additional element $F_4 = -h_2h_4-4q_3\psi_1 H$, when calculating $\{K_2,F_2\}$ and $\{K_3,F_1\}$. The formula for $F_4$ is easily determined from Table \ref{Tab:g1234}, by considering $\{e_3,f_2h_2\}=-h_2 h_4$. The cubic expressions $\{F_i,F_j\}$ can all be written in terms of our linear and quadratic functions. The full set of Poisson relations is given in Table~\ref{Tab:h1-alpha=0:Pij}.
Since $H$ is invariant under the action of $\iota_{ef}$, the Poisson algebra has this symmetry. The action of $\iota_{ef}$ is summarised in Table~\ref{Tab:alpha=0:ief}. Several of the entries in Table~\ref{Tab:h1-alpha=0:Pij} are related through this involution.
\begin{table}[t]\centering
\caption{The action of $\iota_{ef}$ on $H$ and its integrals.}\label{Tab:alpha=0:ief}\vspace{1mm}
\renewcommand\arraystretch{1.26}\begin{tabular}{|c||c||c|c|c||c|c|c||c|}
\hline
 & $H$ &$K_1$ &$K_2$ &$K_3$ &$F_1$ &$F_2$ &$F_3$ &$F_4$ \\[.10cm]\hline\hline
$\iota_{ef}$&$H$ &$-K_1$ &$-\frac12 K_3$ &$-2K_2$ &$-\frac12F_2$ &$-2F_1$ &$F_3$ &$F_4$\\[0.2mm]\hline
\end{tabular}
\end{table}

Adding $H$, we have an 8-dimensional algebra obeying the constraints:
\begin{gather*}
I_1=K_1^2+2K_2K_3+F_3-4\beta H=0,\qquad I_2 = K_1F_4-2K_2F_2+2K_3F_1=0.
\end{gather*}
\begin{Remark}
Both of these are Casimirs of the algebra, with the first extending the Casimir of the isometry algebra and the second being related to ${\cal C}_4^{(13)}$ of (\ref{g13}).
\end{Remark}

\section[Systems with $K=h_4$]{Systems with $\boldsymbol{K=h_4}$}\label{sec:K=h4}

In the case $K=h_4$, we consider the Hamiltonian (\ref{3d-gen}) of the form
\begin{gather}\label{H3d-h4}
H = \varphi(q_1,q_2,q_3)\big(p_1^2+p_2^2+p_3^2\big) = \Phi\big(q_1,q_2^2+q_3^2\big) \big(p_1^2+p_2^2+p_3^2\big),
\end{gather}
which is the most general form commuting with $h_4$. This is also generally invariant with respect to the involution $\iota_{23}$, so we choose our functions $F_i$ to have this invariance.

\subsection[$F_1=e_1^2+\psi_1(q_1,q_2,q_3)H$, $F_2=e_1e_2+\psi_2(q_1,q_2,q_3)H$, $F_3=e_1e_3+\psi_3(q_1,q_2,q_3)H$]{$\boldsymbol{F_1=e_1^2+\psi_1(q_1,q_2,q_3)H}$, $\boldsymbol{F_2=e_1e_2+\psi_2(q_1,q_2,q_3)H}$, \\ $\boldsymbol{F_3=e_1e_3+\psi_3(q_1,q_2,q_3)H}$}\label{sec:h4-F1e12}

With the Hamiltonian (\ref{H3d-h4}), with first integral $K=h_4$, we consider functions of the form
\begin{gather*}
F_1 = e_1^2+\psi_1(q_1,q_2,q_3)H, \quad F_2 = e_1e_2+\psi_2(q_1,q_2,q_3)H,\nonumber\\ F_3 = e_1e_3+\psi_3(q_1,q_2,q_3)H,
\end{gather*}
and require the conditions $\{F_i,H\}=0$ to find
\begin{gather}
 H = \varphi(q_1,q_2,q_3)\big(p_1^2+p_2^2+p_3^2\big) = \frac{1}{\alpha\big(q_1^2+q_2^2+q_3^2\big)+\beta q_1+\gamma} \big(p_1^2+p_2^2+p_3^2\big), \label{h4-H}\\
 \psi_1=-\alpha q_1^2-\beta q_1,\qquad \psi_2=-\left(\alpha q_1+\frac{\beta}{2}\right)q_2,\qquad \psi_3=-\left(\alpha q_1+\frac{\beta}{2}\right)q_3,\nonumber 
\end{gather}
which clearly transform as $(H,F_1,F_2,F_3)\mapsto (H,F_1,F_3,F_2)$ under $\iota_{23}$.

\begin{Remark}[the St\"ackel transform] The Hamiltonian (\ref{h4-H}), can be regarded as the St\"ackel transform of the Euclidean kinetic energy with potential $V=\varphi^{-1}$. In the classification of~\cite{17-3}, this example corresponds to a reduction of $V_{[0]}$, given in \cite[Section~5]{17-3}, by equating $(q_1,q_2,q_3)=(x,y,z)$ and $(\alpha,\beta,\gamma)=(a_4,a_1,a_5)$.
\end{Remark}

This $H$ has first order symmetries
\begin{subequations}\label{h4-Ki}
\begin{gather}\label{h4-K123}
 K_1=h_4,\qquad K_2=\alpha h_2+\beta e_2,\qquad K_3=\alpha h_3+\beta e_3,
\end{gather}
satisfying
\begin{gather}\label{h4-KiKj}
 \{K_1,K_2\}=-4K_3,\qquad \{K_1,K_3\}=4K_2,\qquad \{K_3,K_2\}=\alpha^2 K_1.
\end{gather}
The action of $K_i$ on $F_2$ and $F_3$ introduces three new quadratic elements to our algebra:
\begin{gather}\label{k4-F456}
 F_4=e_2^2-\alpha q_2^2 H, \qquad F_5=e_2e_3-\alpha q_2q_3 H, \qquad F_6=e_3^2-\alpha q_3^2 H,
\end{gather}
\end{subequations}
giving 9 first integrals, which satisfies the Poisson relations of Table~\ref{Tab:h4-Pij}. Adding $H$ gives us a~10-dimensional algebra obeying the constraints:
\begin{gather*}
 I_1 = \alpha K_1F_2-2K_2F_5+2K_3F_4=0,\qquad I_2 = \alpha K_1F_3-2K_2F_6+2K_3F_5=0, \\
 I_3 = 4\alpha K_1F_1-\beta^2 K_1 H-8K_2F_3+8K_3F_2 = 0, \\
 I_4=8F_3F_4-8F_2F_5+K_1K_2H = 0, \qquad I_5 = 8F_3F_5-8F_2F_6+K_1K_3H = 0.
\end{gather*}
\begin{Remark}The first three of these are related to the Casimir ${\cal C}_4^{(he)}$ of~(\ref{cashe}). The elements $K_i$ and $F_i$ are built from the algebra $\mathfrak{g}^{(he)}$ of Section~\ref{conformal-gij}. We also have a {\em linear} relation $F_1+F_4+F_6=\gamma H$, which could be used to eliminate~$F_6$, but this would complicate Table~\ref{Tab:h4-i12}.
\end{Remark}

\begin{table}[t]\centering
\caption{The Poisson algebra of first integrals of (\ref{h4-H}).}\label{Tab:h4-Pij}\vspace{1mm}
\resizebox{\textwidth}{!}{\renewcommand\arraystretch{1.26}\begin{tabular}{|@{\,\,}c@{\,\,}||@{\,\,}c@{\,\,}|@{\,\,}c@{\,\,}|@{\,\,}c@{\,\,}||@{\,\,}c@{\,\,}|@{\,\,}c@{\,\,}|@{\,\,}c@{\,\,}||@{\,\,}c@{\,\,}|@{\,\,}c@{\,\,}|@{\,\,}c@{\,\,}|}
\hline
 &$K_1$ &$K_2$ &$K_3$ &$F_1$ &$F_2$ &$F_3$ &$F_4$ &$F_5$ &$F_6$\\[.10cm]\hline\hline
$K_1$ &$0$ &$-4K_3$ &$4K_2$ &$0$ &$-4F_3$ &$4F_2$ &$-8F_5$ &$4(F_4-F_6)$ &$8F_5$\\[0.2mm]\hline
$K_2$ & &$0$ &$-\alpha^2K_1$ &$4\alpha F_2$ &$2\alpha(F_4-F_1)+\frac12\beta^2H$ &$2\alpha F_5$ &$-4\alpha F_2$ &$-2\alpha F_3$ &$0$\\[0.2mm]\hline
$K_3$ & & &$0$ &$4\alpha F_3$ &$2\alpha F_5$ &$2\alpha(F_6-F_1)+\frac12\beta^2H$ &$0$ &$-2\alpha F_2$ &$-4\alpha F_3$\\[0.2mm]\hline\hline
$F_1$ & & & &$0$ &$-K_2H$ &$-K_3H$ &$0$ &$0$ &$0$\\[0.2mm]\hline
$F_2$ & & & & &$0$ &$\frac14\alpha K_1H$ &$-K_2H$ &$-\frac12K_3H$ &$0$\\[0.2mm]\hline
$F_3$ & & & & & &$0$ &$0$ &$-\frac12K_2H$ &$-K_3H$\\[0.2mm]\hline\hline
$F_4$ & & & & & & &$0$ &$\frac12\alpha K_1H$ &$0$\\[0.2mm]\hline
$F_5$ & & & & & & & &$0$ &$\frac12\alpha K_1H$\\[0.2mm]\hline
$F_6$ & & & & & & & & &$0$ \\[0.2mm]\hline
\end{tabular}}
\end{table}

The Hamiltonian (\ref{h4-H}) is invariant under the action of $\iota_{23}$, as is this algebra (see Table~\ref{Tab:h4-i12}).
\begin{table}[t]\centering
\caption{The action of $\iota_{23}$ on $H$ and its integrals.}\label{Tab:h4-i12}\vspace{1mm}
\renewcommand\arraystretch{1.26}\begin{tabular}{|c||c||c|c|c||c|c|c||c|c|c|}
\hline
 & $H$ &$K_1$ &$K_2$ &$K_3$ &$F_1$ &$F_2$ &$F_3$ &$F_4$ &$F_5$ &$F_6$ \\[.10cm]\hline\hline
$\iota_{23}$&$H$ &$-K_1$ &$K_3$ &$K_2$ &$F_1$ &$F_3$ &$F_2$ &$F_6$ &$F_5$ &$F_4$\\[0.2mm]\hline
\end{tabular}
\end{table}
Several of the entries of Table~\ref{Tab:h4-Pij} (such as $\{K_2,F_2\}$ and $\{K_3,F_3\}$) are related through this involution.

\section{Reductions of the system of Section~\ref{sec:F1=e2f2+}}\label{sec:F1=e2f2+red}

We now consider reductions of the Hamiltonian
\begin{gather*}
H = \frac{q_2^2q_3^2}{\alpha q_2^2+\beta q_3^2+\gamma q_2^2q_3^2}\big(p_1^2+p_2^2+p_3^2\big),
\end{gather*}
associated with the restrictions of Sections \ref{sec:F1=e2f2+:a=0:pb} and \ref{sec:F1=e2f2+:g=0:pb}. We adapt coordinates to the appropriate invariant, rendering it as $P_3$. The 2-dimensional reduction is then $P_3=\mu$, but we keep the label $P_3$ so that we can later transform (in the 3-dimensional context) between these reductions.

In this first section on reductions we give more detailed calculations. The same methods are used in later sections, but we omit the details.

\subsection[Reductions when $\alpha=0$]{Reductions when $\boldsymbol{\alpha=0}$}\label{sec:F1=e2f2+red:alpha=0}

For the case with (\ref{F1=e2f2+:alpha=0:phipsi}) and isometry algebra (\ref{F1=e2f2+:alpha=0:e13h3}) we use respectively $K_2=e_3$ and $K_3=h_3$ to reduce to 2 dimensions.

\subsubsection[Reduction using the isometry $K_2=e_3$]{Reduction using the isometry $\boldsymbol{K_2=e_3}$}

Our original coordinate system $(q_1,q_2,q_3)$ is already adapted to $e_3=p_3=P_3$. Since $\{F_2,e_3\}= \{F_3,e_3\}=0$, we can reduce these integrals in this case, with
\begin{gather}\label{D2+V}
H = \frac{q_2^2}{\beta + \gamma q_2^2} \big(p_1^2+p_2^2\big) + \frac{q_2^2 p_3^2}{\beta + \gamma q_2^2},\qquad F_2= e_2 h_2 - \frac{2 \beta q_1}{q_2^2} H,\qquad
 F_3 = e_2^2-\frac{\beta}{q_2^2} H,
\end{gather}
still satisfying $\{H,e_1\}=0$, $\{F_2,e_1\}=2 F_3$, $\{F_3,e_1\}=0$. The system (\ref{D2+V}) is just the Darboux--Koenigs system of type $D_2$ with the potential of ``Type~D'' in~\cite{03-11}.

By considering linear and quadratic conformal elements (in 3D) that Poisson commute with~$e_3$, we can reduce the $10$-dimensional conformal algebra to the required $6$-dimensional algebra. For example
\begin{gather*}
e_1 ,\ e_2,\ h_2,\ e_1 h_1 - e_3 h_3,\ h_3^2-4 e_1 f_1,\ \mbox{and}\ h_4^2-8 e_2 f_2,
\end{gather*}
give us the standard 6-dimensional subalgebra of conformal symmetries of 2D space:
\begin{gather*}
 T_1=p_1,\qquad T_2=p_2,\qquad T_3 = q_2p_1-q_1p_2,\qquad T_4 = q_1p_1+q_2p_2, \\
 T_5 = \big(q_1^2-q_2^2\big)p_1+2 q_1 q_2 p_2,\qquad T_6 = 2 q_1 q_2 p_1+\big(q_2^2-q_1^2\big)p_2.
\end{gather*}

\begin{Remark}[the reduction of quadratic elements] In the case of quadratic expressions we set $p_3=0$, with only the parts involving $p_1^2$, $p_1p_2$, $p_2^2$ representing a quadratic conformal element of the 2-dimensional kinetic energy. From the 3 linear and 3 quadratic expressions, we can deduce 6 linear elements. For example,
\begin{gather*}
e_1 = p_1, \qquad e_1h_1-e_3h_3 = -2 p_1 (q_1p_1+q_2p_2) -2 q_1 p_3^2
\end{gather*}
give us $T_1$ and $T_4$.
\end{Remark}

The functions $F_2$ and $F_3$ in (\ref{D2+V}) can be written in terms of these, with $T_1$ corresponding to the single Killing vector:
\begin{gather*}
F_2 = -2 T_2 T_3 -2 \beta \frac{q_1}{q_2^2} H , \qquad F_3 = T_2^2 -\frac{\beta}{q_2^2} H,
\end{gather*}
satisfying
\begin{gather*}
\{T_1,F_2\} = -2 F_3,\qquad \{T_1,F_3\} = 0,\qquad \{F_2,F_3\} = 4 T_1 \big(T_1^2+\mu^2 - \gamma H\big).
\end{gather*}

\subsubsection[Reduction using the isometry $K_3=h_3$]{Reduction using the isometry $\boldsymbol{K_3=h_3}$}

We adapt coordinates to $h_3$:
\begin{gather}\label{h3Red:Q123}
Q_1=\sqrt{q_1^2+q_3^2},\qquad Q_2 = q_2,\qquad Q_3 = -\tfrac{1}{2} \arctan\left(\frac{q_1}{q_3}\right) \quad\Rightarrow\quad h_3 = P_3,
\end{gather}
and
\begin{gather}\label{h3RedD2}
H = \frac{Q_2^2}{\beta+\gamma Q_2^2} \left(P_1^2+P_2^2+\frac{P_3^2}{4 Q_1^2}\right),
\end{gather}
which is just the Darboux--Koenigs system of type $D_2$ with the potential of ``Type~B'' (parame\-ter~$b_3$) in~\cite{03-11}.

By considering linear and quadratic conformal elements (in 3D) that Poisson commute with~$h_3$, we can reduce the $10$-dimensional conformal algebra to the required $6$-dimensional algebra. For example
\begin{gather*}
h_1 ,\ e_2,\ f_2,\ e_1^2+ e_3^2,\ 4 f_1^2+f_3^2,\ \mbox{and}\ 2 e_1 h_2+e_3 h_4,
\end{gather*}
give us (setting $P_3=0$) the standard conformal algebra for 2D space:
\begin{gather*}
 T_1=P_1,\qquad T_2=P_2,\qquad T_3 = Q_2P_1-Q_1P_2,\qquad T_4 = Q_1P_1+Q_2P_2, \\
 T_5 = \big(Q_1^2-Q_2^2\big)P_1+2 Q_1 Q_2 P_2,\qquad T_6 = 2 Q_1 Q_2 P_1+\big(Q_2^2-Q_1^2\big)P_2.
\end{gather*}
Furthermore, since $e_1$ and $e_3$ are first integrals, so is
\begin{gather*}
J_2 = e_1^2+e_3^2 = T_1^2+\frac{P_3^2}{4 Q_1^2}.
\end{gather*}
The other independent quadratic integral is $F_1$, which, in these coordinates, takes the form
\begin{gather*}
J_1 = -2 T_1 T_5 + \frac{Q_2^2-Q_1^2}{2 Q_1^2} P_3^2- 2 \big(\beta-\gamma Q_1^2\big) H.
\end{gather*}
From the definition of the constraint $I_1$ in (\ref{F1=e2f2+:alpha=0:I123}), the integral $F_3=-K_1^2-K_2^2+\gamma H = -J_2+\gamma H$, so is not independent.

\begin{Remark}[transformation of potentials]Notice that the change of coordinates~(\ref{h3Red:Q123}) transforms the Hamiltonian of~(\ref{D2+V}) to that of~(\ref{h3RedD2}), {\it preserving} the form of the 2-dimensional met\-ric~$D_2$, but changing the potential.
\end{Remark}

\begin{Remark}[reduction with $\beta = 0$] To get this reduction, we can apply the involution $\iota_{23}$ to the case $\alpha=0$.
\end{Remark}

\subsection[Reductions when $\gamma=0$]{Reductions when $\boldsymbol{\gamma=0}$}\label{sec:F1=e2f2+red:gamma=0}

Now we have
\begin{gather*}
\varphi=\frac{q_2^2q_3^2}{\alpha q_2^2+\beta q_3^2} \qquad\mbox{and}\qquad \psi=-2\beta \frac{q_1^2+q_3^2}{q_2^2},
\end{gather*}
so still have the involutive symmetry $\iota_{23}$ (extended to include $\alpha \leftrightarrow \beta$). We have also gained additional Killing vectors, giving us the {\em isometry algebra} $\mathfrak{g}_1=\langle e_1,h_1,f_1\rangle $. We can therefore reduce with respect to either $h_1$ or $f_1$.

\subsubsection[Reduction using the isometry $K_2=h_1$]{Reduction using the isometry $\boldsymbol{K_2=h_1}$}

We now want to reduce our Hamiltonian (\ref{H3d-e1}) to the submanifold $h_1=\mu$. To do this, we choose canonical coordinates $Q_i, P_i$ so that $h_1=P_3$, which makes $H$ independent of $Q_3$. It is easy to see that
\begin{gather*}
\left\{\frac{q_1}{q_3},h_1\right\}=0,\qquad \left\{\frac{q_2}{q_3},h_1\right\}=0,\qquad \left\{-\tfrac{1}{2}\log q_3,h_1\right\}=1,
\end{gather*}
so clearly we should choose $Q_3 = -\frac{1}{2}\log q_3$ and then some convenient functions of $z_1=\frac{q_1}{q_3}$, $z_2=\frac{q_2}{q_3}$ for $Q_1$, $Q_2$. To obtain a diagonal metric on our reduced space we need the orthogonality relation $\sum\limits_{i,j=1}^3 g^{ij}\pa_iQ_1\pa_jQ_2=0$. We choose $Q_1=\frac{q_1}{q_3}$ and $Q_2 = s\big(\frac{q_1}{q_3},\frac{q_2}{q_3}\big)$, leading to
\begin{gather*}
\big(\big(z_1^2+1\big)\pa_{z_1}+z_1z_2\pa_{z_2}\big)s(z_1,z_2)=0 \quad\Rightarrow\quad s(z_1,z_2) = f\left(\frac{z_2}{\sqrt{z_1^2+1}}\right).
\end{gather*}
With the point transformation
\begin{subequations}\label{Q123H}
\begin{gather}\label{Q123}
Q_1=\frac{q_1}{q_3},\qquad Q_2 = \frac{\sqrt{q_1^2+q_3^2}}{q_2},\qquad Q_3 = -\tfrac{1}{2}\log q_3,
\end{gather}
we find
\begin{gather}\label{HQ}
H = \sum_{i,j=1}^3 \hat g^{ij}P_i P_j,\qquad
 \hat g^{ij} = \frac{1}{\delta} \left(
 \begin{matrix}
 \big(1+Q_1^2\big)^2 & 0 & \frac{1}{2}Q_1 \big(1+Q_1^2\big) \\[1mm]
 0 & Q_2^2 \big(1+Q_2^2\big) & -\frac{1}{2} Q_2 \\[1mm]
 \frac{1}{2}Q_1 \big(1+Q_1^2\big) & -\frac{1}{2} Q_2 & \frac{1}{4}\big(1+Q_1^2\big)
 \end{matrix} \right) ,
\end{gather}
\end{subequations}
where $\delta = \alpha \big(1+Q_1^2\big)+\beta Q_2^2$.

\subsubsection*{The resulting 2D system}

By setting $P_3=\mu$, we can consider our Hamiltonian as $2$-dimensional, with (upper index) metric~$\hat g^{ij}$, given by the $2\times 2$ diagonal matrix in~(\ref{HQ}). However, since we wish to consider this as an embedding into 3 dimensions, it is better to leave $P_3$ in the formulae, remembering that we can always set $P_3=\mu$ if we wish to consider the $2$-dimensional reduction. The quadratic terms~$P_iP_3$ are incorporated in the formulae by ``completing the square'', writing the Hamiltonian as one with electro-magnetic, vector potential:
\begin{gather}\label{HQA}
 H =\sum_{i=1}^2 \hat g^{ii} (P_i-A_i P_3)^2 +h(Q_1,Q_2)P_3^2,
\end{gather}
where
\begin{gather*}
A_1 = \frac{-Q_1}{2\big(1+Q_1^2\big)},\qquad A_2 = \frac{1}{2Q_2\big(1+Q_2^2\big)},\qquad h=\frac{Q_2^2}{4 \big(1+Q_1^2\big) \big(\alpha \big(1+Q_1^2\big)+\beta Q_2^2\big)}.
\end{gather*}

Since, for these particular $A_i$ we have $\{P_1-A_1P_3,P_2-A_2P_3\} = (\pa_1A_2-\pa_2A_1)P_3=0$, we may redefine the momenta (a {\em gauge transformation}) by $\hat P_i = P_i-A_iP_3$, so that~(\ref{HQA}) takes the form $H = \sum\limits_{i=1}^2 \hat g^{ii} \hat P_i^2 +\hat h(Q_1,Q_2)P_3^2$. To incorporate this into the canonical transformation, we must adjust~$Q_3$:
\begin{gather*}
\hat Q_3 = Q_3 + M_1(Q_1)+M_2(Q_2) \qquad\mbox{so that}\qquad \big\{\hat Q_3,P_i-A_i(Q_i) P_3\big\} = 0,
\end{gather*}
leading to $M_1 = -\frac{1}{4} \log \big(1+Q_1^2\big)$, $M_2 = \frac{1}{4} \log\big(\frac{Q_2^2}{1+Q_2^2}\big)$, giving the point transformation (dropping the hats):
\begin{gather*}
Q_1=\frac{q_1}{q_3},\qquad Q_2 = \frac{\sqrt{q_1^2+q_3^2}}{q_2},\qquad Q_3 = -\tfrac{1}{4} \log \big(q_1^2+q_2^2+q_3^2\big).
\end{gather*}

In the original $3$-dimensional setting we had the isometry algebra $\mathfrak{g}_1=\langle e_1,h_1,f_1\rangle $. Since neither of $e_1$, $f_1$ commute with $h_1$, we cannot reduce these to the $2$-dimensional setting. However, since $\{e_1 f_1,h_1\}=0$, we can reduce this {\em quadratic element}, which takes the form
\begin{gather*}
e_1 f_1 = T_1^2 + \left(P_1+\frac{Q_1Q_2P_2}{1+Q_1^2}\right) Q_1P_3, \qquad T_1 = \frac{1}{Q_2} \sqrt{\frac{1+Q_2^2}{1+Q_1^2}} \big(\big(1+Q_1^2\big) P_1+Q_1Q_2P_2\big), \!\!\!
\end{gather*}
where $T_1$ is a first order, first integral of the kinetic energy $H_0 = \sum\limits_{i=1}^2 \hat g^{ii} P_i^2$.

\subsubsection*{Adapting coordinates to $\boldsymbol{T_1}$}

By noting that
\begin{gather*}
y_1 =\frac{Q_2}{\sqrt{1+Q_1^2}},\qquad y_2= \log \left(\frac{Q_1Q_2}{\sqrt{1+Q_1^2}}+\sqrt{1+Q_2^2}\right)
\end{gather*}
satisfy $\{y_1,T_1\}=0$, $\{y_2,T_1\}=1$, we may choose $(y_1,y_2+B(y_1))$ as new coordinates, with $B(y_1)$ chosen so that the resulting metric is diagonal. In this way we find
\begin{gather*}
u = \frac{Q_2}{\sqrt{1+Q_1^2}},\qquad v = \log \left(\frac{Q_1Q_2+\sqrt{\big(1+Q_1^2\big)\big(1+Q_2^2\big)}}{\sqrt{1+Q_1^2+Q_2^2}}\right),
\end{gather*}
in which coordinates $T_1$ and $H$ take the form
\begin{gather*}
T_1=p_v,\qquad H = \frac{1}{\alpha+\beta u^2} \left(u^2\big(1+u^2\big) p_u^2+\frac{u^2}{1+u^2}\, p_v^2+\frac{u^2 \operatorname{sech}^2v\, p_w^2}{4\big(1+u^2\big)}\right).
\end{gather*}
The \looseness=-1 final step is to put $H_0$ into {\em conformal form} $H_0 = \bar \varphi(\bar u) \big(p_{\bar u}^2+p_{\bar v}^2\big)$, in which case, the Laplace--Beltrami operator takes a very simple form: $L_b=\varphi(\bar u,\bar v) \big(\pa_{\bar u}^2\pm \pa_{\bar v}^2\big)$, with {\em no first order terms}. As a consequence, the coordinates $\bar u$ and $\bar v$ satisfy $L_b\bar u=L_b\bar v=0$. Solutions of these equations are
\begin{gather*}
\bar u = \arctan (u),\qquad \bar v = v,
\end{gather*}
which, dropping bars, gives
\begin{gather}\label{D4+V1}
T_1=p_v,\qquad H = \frac{\sin^2(2 u)}{2(\alpha+\beta)+2(\alpha-\beta) \cos(2u)} \big(p_u^2+p_v^2+\tfrac{1}{4}\operatorname{sech}^2v\, p_w^2\big),
\end{gather}
which has the $D_4$ kinetic energy, with potential (equivalent to the ``$b_3$'' part of Case~B in~\cite{03-11}).

We can see that $h_1$ commutes with the 6 functions
\begin{gather*}
e_1 f_1,\ e_2 f_2,\ e_3 f_3,\ h_2,\ h_3,\ \mbox{and}\ h_4,
\end{gather*}
which correspond to the $6$-dimensional conformal algebra of our 2D kinetic energy~$H_0$. In the present coordinates, these take the form
\begin{gather*}
 T_1 = p_v, \qquad\! T_2 = p_u \cosh v \sin u +p_v \sinh v \cos u, \qquad\! T_3 = p_u \cosh v \cos u - p_v \sinh v \sin u, \\
 T_4 = p_u \sinh v \sin u +p_v \cosh v \cos u, \qquad\! T_5 = p_u \sinh v \cos u - p_v \cosh v \sin u, \qquad\! T_6 = p_u.
\end{gather*}
We have seen that both $J_1=F_1$ and $J_2=e_1f_1$ commute with $h_1$ (in the 3D space), so can be reduced to the $2$-dimensional setting. We therefore have that
\begin{gather*}
J_1 = 2 T_2^2 + \beta (\cos 2u-\cosh 2v)\operatorname{sech}^2u\, H - \tfrac{1}{4} \cos^2 u\operatorname{sech}^2 v\, p_w^2 ,\\
J_2 = T_1^2 - \tfrac{1}{4}\tanh^2 v p_w^2 ,
\end{gather*}
commute with $H$.

\subsubsection[Reduction using the isometry $K_3=f_1$]{Reduction using the isometry $\boldsymbol{K_3=f_1}$}

We now want to reduce our Hamiltonian~(\ref{H3d-e1}) to the submanifold $f_1=\mu$, so choose canonical coordinates $Q_i$, $P_i$ so that $f_1=P_3$, which makes $H$ independent of~$Q_3$. It is easy to see that
we should choose $Q_3=\frac{q_1}{q_1^2+q_2^2+q_3^2}$ and then some convenient functions of $z_1=\frac{q_3}{q_2}$ and $z_2=\frac{q_1^2+q_2^2+q_3^2}{q_2}$ for $Q_1$ and $Q_2$. To obtain a diagonal metric on our reduced space we need the orthogonality relation $\sum\limits_{i,j=1}^3 g^{ij}\pa_iQ_1\pa_jQ_2=0$, which leads us to
\begin{gather*}
 Q_1=\frac{q_3}{q_2},\qquad Q_2=\frac{q_1^2+q_2^2+q_3^2}{\sqrt{q_2^2+q_3^2}},\qquad Q_3=\frac{q_1}{q_1^2+q_2^2+q_3^2},
\end{gather*}
and
\begin{gather*}
 H = \frac{Q_1^2}{\alpha+\beta Q_1^2}\left(\big(Q_1^2+1\big)P_1^2+\frac{Q_2^2}{Q_1^2+1}P_2^2+\frac{P_3^2}{Q_2^2\big(Q_1^2+1\big)}\right).
\end{gather*}
From the Casimir ${\cal C}=e_1f_1+\frac{1}{4} h_1^2$, of our isometry algebra, we obtain
\begin{gather}\label{cas-f1-red}
{\cal C} = T_1^2+\frac{P_3^2}{Q_2^2},
\end{gather}
where $T_1 = Q_2 P_2$ corresponds to a Killing vector on the reduced space.

\subsubsection*{Adapting coordinates to $\boldsymbol{T_1}$}

We have
\begin{gather*}
\{Q_1,T_1\}=\{Q_3,T_1\}=0 \qquad\mbox{and}\qquad \{\log(Q_2),T_1\}=1,
\end{gather*}
so must build coordinates out of these, such that the kinetic energy $H_0$ is transformed into {\em conformal form} $H_0 = \varphi(u) \big(p_{u}^2+p_{v}^2\big)$. This is achieved by the choice
\begin{gather*}
u = \arctan(Q_1),\qquad v = \log(Q_2), \qquad w = Q_3 \qquad\Rightarrow\qquad T_1 = p_v
\end{gather*}
and
\begin{gather}\label{Hu-f1}
H = \frac{\sin^2(2 u)}{2(\alpha+\beta)+2(\alpha-\beta) \cos(2u)} \big(p_u^2+p_v^2+{\rm e}^{-2v} p_w^2\big),
\end{gather}
which has the $D_4$ kinetic energy, with potential (equivalent to the~``$a_1$'' part of Case~A in~\cite{03-11}).

\subsubsection*{The 6-dimensional conformal algebra}

We can identify 3 linear and 3 quadratic conformal elements in the 3D space, which commute with $f_1$, so can be reduced to the 2-dimensional setting:
\begin{gather*}
e_1f_1+\tfrac{1}{4} h_1^2,\ f_2,\ f_3,\ h_4,\ e_2f_2-\tfrac{1}{2} h_2^2,\ e_3f_3+\tfrac{1}{2} h_3^2,
\end{gather*}
leading to
\begin{gather*}
 T_1 = p_v,\qquad T_2 = {\rm e}^v (\sin u p_u+\cos u p_v),\qquad T_3 = {\rm e}^v (\cos u p_u-\sin u p_v),\nonumber\\
T_4 = p_u,\qquad T_5 = {\rm e}^{-v} (\sin u p_u-\cos u p_v),\qquad T_6 = {\rm e}^{-v} (\cos u p_u+\sin u p_v),
\end{gather*}
with $T_1$ being the sole isometry.

The Hamiltonian (\ref{Hu-f1}) has 2 independent quadratic integrals, derived from the Casi\-mir~(\ref{cas-f1-red}) and the former integral~$F_5$ (see~(\ref{F45})), which can be written in terms of the above conformal elements:
\begin{gather*}
J_1 = T_1^2+{\rm e}^{-2 v}\, p_w^2,\qquad J_2 = 4 T_2^2- 4 \beta {\rm e}^{2 v} \sec^2 u\, H.
\end{gather*}

\subsubsection{The relationship between these reductions}

If we consider the $h_1$ reduction to have coordinates $(\bar u,\bar v,\bar w)$, then combining the transformations, we obtain
\begin{gather*}
u = \bar u,\qquad v = 2 \bar w +\log(\cosh \bar v),\qquad w = {\rm e}^{2 \bar w} \tanh\bar v,
\end{gather*}
and the Hamiltonian (\ref{Hu-f1}) is transformed onto that of~(\ref{D4+V1}) (with bars). This transformation preserves the form of the 2-dimensional kinetic energy, but changes the potential, thus relating two of the potentials obtained in~\cite{03-11}.

\section{Reductions of the system of Section \ref{sec:FGe1h2}}\label{sec:FGe1h2-red}

We saw in Section \ref{sec:FGe1h2} that the Hamiltonian
\begin{gather*}
 H=\frac{1}{\alpha q_2+\beta q_3+\gamma}\big(p_1^2+p_2^2+p_3^2\big)
\end{gather*}
admitted the $3$-dimensional isometry algebra (\ref{KiFGe1h2}), with Poisson relations (\ref{KijFGe1h2}). In this section we reduce this Hamiltonian to a 2D space in two different ways, using the isometries $K_2$ and $K_3$ respectively.

\subsection[Reduction using the isometry $K_2=\beta e_2-\alpha e_3$]{Reduction using the isometry $\boldsymbol{K_2=\beta e_2-\alpha e_3}$}\label{sec:FGe1h2-redK2}

We see that
\begin{gather*}
\{\alpha q_2+\beta q_3,K_2\}=\{q_1,K_2\}=0 \qquad\mbox{and}\qquad \left\{\frac{q_2}{\beta},K_2\right\}=1,
\end{gather*}
and we need to choose new coordinates $(Q_1,Q_2,Q_3)$, such that $K_2=P_3$ (so the new $H$ is independent of~$Q_3$) and $H$ takes the form
\begin{gather*}
H = \tilde\varphi(Q_1,Q_2) \big(P_1^2+P_2^2\big) +\tilde V(Q_1,Q_2) P_3^2,
\end{gather*}
which can be achieved in stages as described in Section~\ref{sec:F1=e2f2+red} or by the single canonical transformation generated by
\begin{gather}\label{SredK2}
S = \frac{(\alpha q_2+\beta q_3+\gamma)}{\big(\alpha^2+\beta^2\big)^{\frac{1}{3}}}\ P_1 + \big(\alpha^2+\beta^2\big)^{\frac{1}{6}} q_1 P_2 +
 \left(\frac{q_2}{\beta}- \frac{\alpha (\alpha q_2+\beta q_3+\gamma)}{\beta\big(\alpha^2+\beta^2\big)}\right) P_3,
\end{gather}
which transforms the Hamiltonian to
\begin{gather}\label{3dD1-r1}
 H=\frac{1}{Q_1} \big(P_1^2+P_2^2\big)+\frac{P_3^2}{\big(\alpha^2+\beta^2\big)^{\frac{4}{3}}Q_1},
\end{gather}
so the entire Hamiltonian (not just the kinetic energy) has inherited the symmetry $K_1=\big(\alpha^2+\beta^2\big)^{\frac{1}{6}} P_2$, which is a result of the property $\{K_1,K_2\}=0$.

Since $P_3$ is a first integral, we can reduce to the $2$-dimensional space with $P_3=\mu$. Again, we can derive the $6$-dimensional conformal algebra from our original $10$-dimensional algebra~(\ref{g1234}), by considering {\em linear} and {\em quadratic} conformal elements which {\em commute} with~$K_2$. The 3 linear elements
\begin{gather*}
\alpha e_2+\beta e_3,\ e_1,\ \alpha h_2+\beta h_3\qquad\text{give us}\qquad T_1=P_1,\qquad T_2 = P_2,\qquad T_3 = Q_2 P_1-Q_1 P_2,
\end{gather*}
while the 3 quadratic elements
\begin{gather*}
\alpha K_1h_1+ h_3 K_2,\qquad 8(\alpha e_2+\beta e_3)(\alpha f_2+\beta f_3) - \big(\alpha^2+\beta^2\big) h_4^2,\qquad 4 \big(\alpha^2+\beta^2\big) e_1f_1-K_3^2,
\end{gather*}
factorise (when $P_3=0$) to give the 3 remaining conformal elements:
\begin{gather*}
T_4 = Q_1 P_1+Q_2 P_2,\qquad T_5 = \big(Q_1^2-Q_2^2\big)P_1+2 Q_1 Q_2 P_2,\\ T_6 = 2 Q_1 Q_2 P_1+\big(Q_2^2-Q_1^2\big)P_2.
\end{gather*}
The $2$-dimensional Hamiltonian (\ref{3dD1-r1}) (with $P_3=\mu$) is the $D_1$ kinetic energy, with potential (equivalent to the ``$b_2$'' part of Case~1 in~\cite{02-6}).

\subsubsection{The quadratic integrals}

We consider the 2 functions (\ref{Fe1h2}) and see that $\{F_2,K_2\}=0$, but $\{F_1,K_2\}\neq 0$. However, it is easy to see that $\big\{\big(\alpha^2+\beta^2\big) F_1-\beta K_1 K_3,K_2\big\}=\{K_1,K_2\}=0$. Noting that
\begin{gather*}
\big(\alpha^2+\beta^2\big) F_1-\beta K_1 K_3 = \alpha K_1 (\alpha h_2+\beta h_3)-\tfrac{1}{2} \alpha \big(\alpha^2+\beta^2\big) q_1^2 H,
\end{gather*}
and that $\alpha h_2+\beta h_3\mapsto 2 \gamma \big(\alpha^2+\beta^2\big)^{\frac{1}{6}} T_2+2 \big(\alpha^2+\beta^2\big)^{\frac{1}{2}}T_3$ (under the action of the transformation generated by~(\ref{SredK2})) this integral reduces to
\begin{gather*}
J_1 = T_2 T_3 - \tfrac{1}{4} Q_2^2 H = P_2 (Q_2 P_1-Q_1 P_2) - \tfrac{1}{4} Q_2^2 H.
\end{gather*}
The function $F_2$ is more straightforward and reduces to
\begin{gather*}
J_2 = T_1 T_2 -\frac{Q_2}{2} H = P_1 P_2-\frac{Q_2}{2} H.
\end{gather*}
In both these formulae, $H$ is the full (\ref{3dD1-r1}) (with potential) and $\{J_i,H\}=0$, for each $i$, with $\{J_1,P_2\}=4 J_2$.

\subsection[Reduction using the isometry $K_3=\beta h_2-\alpha h_3$]{Reduction using the isometry $\boldsymbol{K_3=\beta h_2-\alpha h_3}$}\label{sec:FGe1h2-redK3}

We see that
\begin{gather*}
\{\alpha q_2+\beta q_3,K_3\}=\big\{q_1^2+q_2^2+q_3^2,K_3\big\}=0 \qquad\mbox{and}\\
\left\{\frac{1}{2\sqrt{\alpha^2+\beta^2}} \tan^{-1}\left(\frac{\beta q_2-\alpha q_3}{q_1\sqrt{\alpha^2+\beta^2}}\right),K_3\right\}=1,
\end{gather*}
and we again need to choose new coordinates $(Q_1,Q_2,Q_3)$, such that $K_2=P_3$ (so the new $H$ is independent of~$Q_3$) and $H$ takes the form
\begin{gather*}
H = \tilde\varphi(Q_1,Q_2) \big(P_1^2+P_2^2\big) +\tilde V(Q_1,Q_2) P_3^2,
\end{gather*}
which can be achieved in stages as described in Section~\ref{sec:F1=e2f2+red} or by the single canonical transformation generated by
\begin{gather*}
S = \frac{\alpha q_2+\beta q_3+\gamma}{\big(\alpha^2+\beta^2\big)^{\frac{1}{3}}} P_1 + \frac{\big(\alpha^2+\beta^2\big) q_1^2+(\beta q_2-\alpha q_3)^2}{\big(\alpha^2+\beta^2\big)^{\frac{1}{3}}} P_2 \nonumber\\
\hphantom{S = }{} + \frac{1}{2\sqrt{\alpha^2+\beta^2}} \tan^{-1}\left(\frac{\beta q_2-\alpha q_3}{q_1\sqrt{\alpha^2+\beta^2}}\right) P_3,
\end{gather*}
which transforms the Hamiltonian to
\begin{gather}\label{3dD1-r2}
 H=\frac{1}{Q_1} \big(P_1^2+P_2^2\big)+\frac{P_3^2}{4\big(\alpha^2+\beta^2\big)Q_1Q_2^2},
\end{gather}
so \looseness=-1 this Hamiltonian has {\em not} inherited the symmetry $K_1$, but the {\em kinetic energy} has acquired the symmetry~$P_2$, which is the reduction of~$K_1$ to the 2D space, corresponding to the level set $P_3=\mu$.

Again, we can derive the $6$-dimensional conformal algebra from our original $10$-dimensional algebra~(\ref{g1234}), by considering {\em linear} and {\em quadratic} conformal elements which {\em commute} with $K_3$. The 3 linear elements
\begin{gather*}
\alpha e_2+\beta e_3,\ h_1,\ \alpha f_2+\beta f_3\qquad\text{give us}\\
 T_1=P_1,\qquad T_4 = Q_1 P_1+Q_2 P_2,\qquad T_5 = \big(Q_1^2-Q_2^2\big)P_1+2 Q_1 Q_2 P_2,
\end{gather*}
while the 3 quadratic elements
\begin{gather*}
4\big(h_2^2+ h_3^2\big)+h_4^2,\qquad K_2^2 +\big(\alpha^2+\beta^2\big) K_1^2, \qquad 4 \big(\alpha^2+\beta^2\big) f_1^2 (\beta f_2-\alpha f_3)^2,
\end{gather*}
simplify (when $P_3=0$) to give the 3 remaining conformal elements:
\begin{gather*}
T_2 = P_2,\qquad T_3 = Q_2 P_1-Q_1 P_2,\qquad T_6 = 2 Q_1 Q_2 P_1+\big(Q_2^2-Q_1^2\big)P_2.
\end{gather*}
The $2$-dimensional Hamiltonian (\ref{3dD1-r2}) (with $P_3=\mu$) is the $D_1$ kinetic energy, with potential (equivalent to the ``$b_3$'' part of Case~1 in~\cite{02-6}).

\subsection{The relationship between these reductions}\label{sec:FGe1h2-redK2K3}

For this section we change notation for the reduction of Section~\ref{sec:FGe1h2-redK2}, writing the 3D coordinates as $(u,v,w)$:
\begin{gather}
 H_u=\frac{1}{u} \big(p_u^2+p_v^2\big)+\frac{p_w^2}{\big(\alpha^2+\beta^2\big)^{\frac{4}{3}}u},\qquad J_{1u} = p_v (v p_u-u p_v) - \tfrac{1}{4} v^2 H_u,\nonumber\\
 J_{2u} = p_u p_v-\frac{v}{2} H_u.\label{HJu}
\end{gather}
Composing the two point transformations $(u,v,w)\mapsto (q_1,q_2,q_3) \mapsto (Q_1,Q_2,Q_3)$, we obtain
\begin{gather*}
u = Q_1,\qquad v = Q_2 \cos\big(2\sqrt{\alpha^2+\beta^2} Q_3\big),\\
 w = \frac1{\big(\alpha^2+\beta^2\big)^\frac{2}{3}} Q_2 \sin\big(2\sqrt{\alpha^2+\beta^2} Q_3\big)-\frac{\alpha}{\beta\big(\alpha^2+\beta^2\big)}.
\end{gather*}
Under the corresponding canonical transformation of the {\em full} $6$-dimensional phase space, the Hamiltonian $H_u$ is transformed to $H_Q$, given by (\ref{3dD1-r2}). This means that it preserves the form of the {\em kinetic energy} (associated with $D_1$) and changes the~``$b_2$'' part of the potential of Case~1 in~\cite{02-6} to the ``$b_3$'' part.

Any other integral $F_u$ is transformed to
\begin{gather*}
F_Q = F_Q^{(0)}+F_Q^{(c)} \cos(k Q_3)+F_Q^{(s)} \sin(k Q_3), \qquad\mbox{for some}\quad k.
\end{gather*}
In the full $6$-dimensional phase space, $\{H_Q,F_Q\}=0$ (noting that the $P_3^2$ term in $H_Q$ will interact with $Q_3$), but in the $2$-dimensional reduction (with $P_3=\mu$), we have $\big\{H_Q,F_Q^{(0)}\big\}=0$.

Carrying this out for $J_{1u}$, $J_{2u}$, we find
\begin{gather*}
J_{1Q}^{(0)} = P_2 (Q_2 P_1-Q_1 P_2) - \frac{Q_1 P_3^2}{4 \big(\alpha^2+\beta^2\big) Q_2^2}-\tfrac{1}{4} Q_2^2 H_Q,\qquad J_{2Q}^{(0)} = 0.
\end{gather*}
The function $J_{2u}^2$ would lead to something nontrivial, but it would be quartic. However, we can use the Casimir (\ref{casKFGe1h2}) of the original isometry algebra:
\begin{gather*}
{\cal C}_K = \big(\alpha^2+\beta^2\big)^{\frac{4}{3}}p_v^2+p_w^2 = \big(\alpha^2+\beta^2\big)^{\frac{4}{3}} \left(P_2^2+\frac{P_3^2}{4 \big(\alpha^2+\beta^2\big) Q_2^2}\right).
\end{gather*}

We have thus found that the system (\ref{HJu}) (together with combinations of its {\em linear} integrals) can be transformed onto the system
\begin{gather*}
 H_Q=\frac{1}{Q_1}\,\big(P_1^2+P_2^2\big)+\frac{P_3^2}{4\big(\alpha^2+\beta^2\big)Q_1Q_2^2},\nonumber\\
 R_{1Q} = P_2 (Q_2 P_1-Q_1 P_2) - \frac{Q_1 P_3^2}{4 \big(\alpha^2+\beta^2\big) Q_2^2}-\frac{1}{4} Q_2^2 H_Q,\nonumber\\
 R_{2Q} = P_2^2+\frac{P_3^2}{4 \big(\alpha^2+\beta^2\big) Q_2^2}, 
\end{gather*}
corresponding to two particular parametric reductions of the potential of Case~1 in~\cite{02-6}.

\section{Reductions of the system of Section \ref{sec:F1e2h2}}\label{sec:F1e2h2-red}

We now consider reductions of the Hamiltonian
\begin{gather*}
H = \frac{q_2^2q_3^2\sqrt{q_1^2+q_2^2}}{(\alpha q_2^2+\beta q_3^2)\sqrt{q_1^2+q_2^2}+\gamma q_1q_3^2}\big(p_1^2+p_2^2+p_3^2\big).
\end{gather*}
We found that we had three cases, corresponding to $\alpha=0$ or $\beta=0$ or $\gamma=0$. However, $\gamma=0$ is just the Hamiltonian discussed in Sections~\ref{sec:F1=e2f2+:g=0:pb} and~\ref{sec:F1=e2f2+red:gamma=0}, so we are left with just the first two. Our first example is the case with $\alpha=0$, which has a $3$-dimensional isometry algebra and reduces to a Hamiltonian with Darboux--Koenigs metric~$D_4$. However, in Section \ref{F1=e2h2+:beta=0-red}, we consider the case $\beta=0$, which has only {\em one} linear integral $h_1$, with the consequence that the reduced metric has {\it no} Killing vectors at all. Nevertheless, the reduced Hamiltonian commutes with a pair of functions, one quadratic and the other quartic, so is a maximally super-integrable system which is {\em outside} the Darboux--Koenigs classification.

\subsection[The case when $\alpha=0$]{The case when $\boldsymbol{\alpha=0}$}\label{F1=e2h2+:alpha=0-red}

We now consider the Hamiltonian
\begin{gather*}
H=\frac{q_2^2\sqrt{q_1^2+q_2^2}}{\beta \sqrt{q_1^2+q_2^2}+\gamma q_1}\big(p_1^2+p_2^2+p_3^2\big),
\end{gather*}
which has the isometry algebra $K_1 = h_1$, $K_2=e_3$, $K_3=f_3$, satisfying
\begin{gather*}
 \{K_1,K_2\}=-2 K_2,\qquad \{K_1,K_3\}=2 K_3,\qquad \{K_2,K_3\}=-2 K_1.
\end{gather*}
We can therefore reduce with respect to either $K_1$ or $K_3$. Reduction with respect to $K_2$ would be equivalent to that with respect to $K_3$ through the action of the involution~$\iota_{ef}$.

\subsubsection[Reduction using the isometry $K_1$]{Reduction using the isometry $\boldsymbol{K_1}$}

Since $K_1=h_1$, we can again use the transformation (\ref{Q123}):
\begin{gather*}
Q_1=\frac{q_1}{q_3},\qquad Q_2 = \frac{\sqrt{q_1^2+q_3^2}}{q_2},\qquad Q_3 = -\tfrac{1}{2}\log q_3,
\end{gather*}
this time giving
 \begin{gather*}
H = \frac{Q_1^2}{\beta \big(1 + Q_1^2\big) + \gamma \sqrt{1 + Q_1^2}} \left(\big(1 + Q_1^2\big)^2\left(P_1 + \frac{Q_1P_3}{2\big(1 + Q_1^2\big)}\right)^2 \right.\nonumber\\
 \left. \hphantom{H =}{} +\big(1 + Q_2^2\big)\left(P_2 + \frac{Q_2P_3}{2\big(1 + Q_2^2\big)}\right)^2 + \frac{P_3^2}{4(1 + Q_2^2)}\right). 
\end{gather*}
Here we have already ``completed the square'' to give the vector potential
\begin{gather*}
(A_1,A_2)=\left(-\frac{Q_1}{2\big(1+Q_1^2\big)},-\frac{Q_2}{2\big(1+Q_2^2\big)}\right),
\end{gather*}
satisfying
\begin{gather*}
\{P_1-A_1P_3,P_2-A_2P_3\}=(\pa_1A_2-\pa_2A_1)P_3=0,
\end{gather*}
so we can use these to define new momenta $\big(\tilde P_1,\tilde P_2\big)$, but must adjust the definition of~$Q_3$ in order to keep the transformation canonical:
\begin{gather*}
\{Q_3+M_1(Q_1)+M_2(Q_2),P_i-A_i P_3\}=0 \qquad\Rightarrow\qquad M_1'(Q_1)=A_1,\qquad M_2'(Q_2)=A_2,
\end{gather*}
leading to the point transformation
\begin{gather*}
Q_1=\frac{q_1}{q_3},\qquad Q_2 = \frac{\sqrt{q_1^2+q_3^2}}{q_2},\qquad Q_3 = -\tfrac{1}{4}\log \big(q_1^2+q_2^2+q_3^2\big),
\end{gather*}
now giving
\begin{gather*} 
H = \frac{Q_1^2}{\beta \big(1+Q_1^2\big)+\gamma \sqrt{1+Q_1^2}} \left(\big(1+Q_1^2\big)^2P_1^2 +\big(1+Q_2^2\big)P_2^2+\frac{P_3^2}{4\big(1+Q_2^2\big)}\right).
\end{gather*}
Since $\{K_2 K_3,K_1\}=0$, this quantity defines a Killing vector for the 2-dimensional metric, associated with this Hamiltonian:
\begin{gather*}
K_2 K_3 = 2\left(T_1^2 -\frac{Q_2^2P_3^2}{4\big(1+Q_2^2\big)}\right) , \quad\mbox{with}\qquad T_1 = \sqrt{\big(1+Q_2^2\big)} P_2 ,
\end{gather*}
where $T_1$ is a first order, first integral of the 2D kinetic energy $H_0$.

We choose coordinates $(u,v,w)$ so that $T_1=p_v$ and $H_0=\tilde \varphi(u) \big(p_u^2+p_v^2\big)$:
\begin{gather*}
Q_1 = \tan u,\qquad Q_2 = \sinh v,\qquad Q_3=w,
\end{gather*}
giving
\begin{gather} \label{h1-Hu}
H = \frac{\sin^2 u}{\beta+\gamma \cos u} \big(p_u^2+p_v^2+\tfrac{1}{4}\operatorname{sech}^2 v\, p_w^2\big),
\end{gather}
which has the $D_4$ kinetic energy, with potential (equivalent to the ``$b_3$'' part of Case~B in~\cite{03-11}).

\subsubsection*{The conformal algebra and quadratic integrals}

We can see that $K_1=h_1$ commutes with the 6 functions
\begin{gather*}
e_3 f_3,\;\;\; e_1 f_1,\;\;\; e_2 f_2,\;\;\; h_2,\;\;\; h_3,\;\;\; \mbox{and}\;\;\; h_4,
\end{gather*}
which correspond to the $6$-dimensional conformal algebra of our 2D kinetic energy $H_0$. In the present coordinates, these take the form
\begin{gather*}
 T_1 = p_v, \qquad\! T_2 = p_u \cosh v \sin u +p_v \sinh v \cos u, \qquad\! T_3 = p_u \cosh v \cos u - p_v \sinh v \sin u, \\
 T_4 = p_u,\qquad\! T_5 = p_u \sinh v \sin u +p_v \cosh v \cos u, \qquad\! T_6 = p_u \sinh v \cos u - p_v \cosh v \sin u.
\end{gather*}
The element $T_1$ is an {\em invariant} of the kinetic energy, but the full expression for $K_2K_3=e_3f_3$ is a quadratic integral for~$H$:
\begin{gather*}
J_1 = \tfrac{1}{2} e_3 f_3 = T_1^2-\tfrac{1}{4} \tanh^2 v\, p_w^2.
\end{gather*}
We also have that $\{F_3,K_1\}=\{F_4,K_1\}=0$ (see Table~\ref{Tab:h1-alpha=0:Pij}), but $F_3 = 4 J_1-4 \beta H+p_w^2$, so is not independent. However~$F_4$ is independent and takes the form
\begin{gather*} 
J_2 = F_4 = - 8 T_4T_6 +2 (4\beta \cos u +\gamma (3 +\cos 2u)) \csc^2 u \sinh v \, H,
\end{gather*}
satisfying
\begin{gather*}
\{J_1,J_2\} = 16 T_1T_3T_4-4 (4\beta \cos u +\gamma (3 +\cos 2u)) \cosh v \csc^2 u\, T_1 H\nonumber\\
\hphantom{\{J_1,J_2\} =}{} - 4 \operatorname{sech} v \tanh v \sin u\, T_4 p_w^2.
\end{gather*}

\subsubsection[Reduction using the isometry $K_3=f_3$]{Reduction using the isometry $\boldsymbol{K_3=f_3}$}

With $K_3=f_3$, it is again possible to choose coordinates such that $K_3=P_3$, and to set the coefficient of $P_1 P_2$ to be zero:
\begin{gather*}
Q_1=\frac{q_2}{q_1},\qquad Q_2 = \frac{q_1^2+q_2^2+q_3^2}{\sqrt{q_1^2+q_2^2}},\qquad Q_3 = \frac{q_3}{2 \big(q_1^2+q_2^2+q_3^2\big)},
\end{gather*}
giving
\begin{gather*} 
H = \frac{Q_1^2}{\beta \big(1+Q_1^2\big)+\gamma \sqrt{1+Q_1^2}} \left(\big(1+Q_1^2\big)^2 P_1^2 +Q_2^2 P_2^2 +\frac{P_3^2}{4 Q_2^2}\right).
\end{gather*}
A further change of coordinates
\begin{gather*}
Q_1=\tan u,\qquad Q_2 = {\rm e}^v,\qquad Q_3 = w,
\end{gather*}
gives
\begin{gather} \label{h1f3-Hu}
H = \frac{\sin^2 u}{\beta +\gamma \cos u} \big(p_u^2+p_v^2 +\tfrac{1}{4} {\rm e}^{-2v} p_w^2\big),
\end{gather}
which has the $D_4$ kinetic energy, with potential (equivalent to the ``$a_1$'' part of Case~A in~\cite{03-11}).

\subsubsection*{The conformal algebra and quadratic integrals}

We can see that $K_3=f_3$ commutes with the 6 functions
\begin{gather*}
2 e_3 f_3+h_1^2,\ 4 e_1 f_1-h_3^2,\ 8 e_2 f_2-h_4^2,\ h_2,\ f_1,\ \mbox{and}\ f_2,
\end{gather*}
which correspond to the $6$-dimensional conformal algebra of our 2D kinetic energy $H_0$. In the present coordinates, these take the form
\begin{gather*}
T_1 = p_v, \qquad T_2 = {\rm e}^v (p_u \sin u +p_v \cos u), \qquad T_3 = {\rm e}^v (p_u \cos u - p_v \sin u), \\
T_4 = p_u,\qquad T_5 = {\rm e}^{-v} (p_u \sin u -p_v \cos u), \qquad T_6 = {\rm e}^{-v} (p_u \cos u + p_v \sin u).
\end{gather*}
The element $T_1$ is an {\em invariant} of the kinetic energy, but the full expression for $2 K_2K_3-K_1^2=2 e_3 f_3+h_1^2$ (the Casimir of the isometry algebra) is a quadratic integral for $H$:
\begin{gather*}
J_1 = \tfrac{1}{4} \big(2 K_2K_3-K_1^2\big) = T_1^2+\tfrac{1}{4} {\rm e}^{-2 v} p_w^2.
\end{gather*}
We also have that $\{F_2,K_3\}=\{F_3,K_3\}=0$ (see Table~\ref{Tab:h1-alpha=0:Pij}), but $F_3 = 4 (J_1- \beta H)$, so is not independent. However~$F_2$ is independent and takes the form
\begin{gather*} 
J_2 = F_2 = - 4 T_3T_4 - (4\beta \cos u +\gamma (3 +\cos 2u)) \csc^2 u \, {\rm e}^v H.
\end{gather*}

\subsubsection{The relationship between these reductions}

If we consider the $K_1$ reduction to have coordinates $(\bar u,\bar v,\bar w)$, then combining the transformations, we obtain
\begin{gather*}
u = \bar u,\qquad v = -2 \bar w +\log(\cosh \bar v),\qquad w = \tfrac{1}{2} {\rm e}^{2 \bar w} \tanh\bar v,
\end{gather*}
and the Hamiltonian (\ref{h1f3-Hu}) is transformed onto that of (\ref{h1-Hu}) (with bars). This transformation preserves the form of the 2-dimensional kinetic energy, but changes the potential, thus relating two of the potentials obtained in \cite{03-11}.

\subsection[The case when $\beta=0$]{The case when $\boldsymbol{\beta=0}$}\label{F1=e2h2+:beta=0-red}

We now consider the Hamiltonian
\begin{gather}\label{F1=e2h2+:beta=0:H}
H=\frac{q_2^2q_3^2\sqrt{q_1^2+q_2^2}}{\alpha q_2^2 \sqrt{q_1^2+q_2^2}+\gamma q_1q_3^2}\big(p_1^2+p_2^2+p_3^2\big),
\end{gather}
which has the {\em single} isometry $K_1 = h_1$. We therefore reduce with respect to this by making an appropriate choice of coordinates~$Q_i$, which would generally give ``vector potential'' terms, which again turn out to be gauge equivalent to zero. We can again incorporate this gauge transformation into the definition of~$Q_3$, to obtain
\begin{gather*}
 Q_1=\frac{q_2}{q_1},\qquad Q_2=\frac{q_3}{\sqrt{q_1^2+q_2^2}},\qquad Q_3=-\tfrac{1}{4} \log \big(q_1^2+q_2^2+q_3^2\big),
\end{gather*}
giving
\begin{gather*}
H=\frac{Q_1^2Q_2^2}{\alpha Q_1^2+\gamma Q_2^2\sqrt{Q_1^2+1}} \left(\big(Q_1^2+1\big)^2 P_1^2+\big(Q_2^2+1\big) P_2^2 +\frac{P_3^2}{4\big(Q_2^2+1\big)}\right).
\end{gather*}
The final transformation puts the kinetic energy into the form $H_0=\tilde{\varphi}(u,v)\big(p_u^2+p_v^2\big)$:
\begin{gather*}
 Q_1 = \tan u,\qquad Q_2 = \sinh v, \qquad Q_3 = w,
\end{gather*}
giving
\begin{gather}\label{beta=0:Huv}
H=\frac{\sin^2{u} \sinh^2{v}}{\alpha\sin^2{u}+\gamma\cos{u}\,\sinh^2{v}}\, \left(p_u^2+p_v^2+\frac{1}{4}\operatorname{sech}^2 v\, p_w^2\right).
\end{gather}
The corresponding metric has no Killing vectors, so we cannot reduce further. However, we can construct the $6$-dimensional conformal algebra from the linear and quadratic invariants of~$h_1$:
\begin{gather*}
h_2,\ h_3,\ h_4,\ e_1f_1,\ e_2f_2,\ e_3f_3,
\end{gather*}
giving
\begin{gather*}
 T_1=p_u,\qquad T_2=\sin{u}\sinh{v}p_u+\cos{u}\cosh{v}p_v,\qquad T_3=\cos{u}\sinh{v}p_u-\sin{u}\cosh{v}p_v, \\
 T_4=\sin{u}\cosh{v}p_u+\cos{u}\sinh{v}p_v,\qquad T_5=\cos{u}\cosh{v}p_u-\sin{u}\sinh{v}p_v,\qquad T_6=p_v.
\end{gather*}
We see from (\ref{PBFK}) that $h_1$ commutes with $F_3$ and $F_1F_2$, which therefore give us a quadratic and a quartic integral for~$H$, which can be written in terms of the conformal algebra:
\begin{gather*}
J_1 = T_1^2 -\gamma \cot u \operatorname{cosec} u\, H, \\ 
J_2 = \big(2 T_1T_5 -\gamma \cosh v\operatorname{cosec}^2u \big(1+\cos^2 u\big) H\big)^2-T_1^2 p_w^2\operatorname{sech}^2 v\sin^2 u. 
\end{gather*}
We thus find that the Hamiltonian~(\ref{beta=0:Huv}) is maximally super-integrable, but that it falls out of the Darboux--Koenigs classification, since it has {\it no} first order integral, together with one quadratic and one {\it quartic} integral.

\section{Reductions of the system of Section \ref{sec:h4-F1e12}}\label{sec:h4-F1e12-red}

We now consider the Hamiltonian
\begin{gather*}
H = \frac{1}{\alpha\big(q_1^2+q_2^2+q_3^2\big)+\beta q_1+\gamma}\big(p_1^2+p_2^2+p_3^2\big),
\end{gather*}
with symmetry algebra $\langle K_1,K_2,K_3\rangle $, defined by (\ref{h4-K123}).

We can therefore reduce with respect to either $K_1$ or $K_2$. Reduction with respect to~$K_3$ would be equivalent to that with respect to~$K_2$ through the action of the involution~$\iota_{23}$.

\subsection[Reduction using the isometry $K_1=h_4$]{Reduction using the isometry $\boldsymbol{K_1=h_4}$}

To transform $K_1=h_4$ to $P_3$ and to eliminate the $P_1 P_2$ term, we choose coordinates
\begin{gather*}
Q_1=q_1,\qquad Q_2=\sqrt{q_2^2+q_3^2},\qquad Q_3=\tfrac{1}{4} \arctan\left(\frac{q_2}{q_3}\right),
\end{gather*}
giving
\begin{gather} \label{h4-HQ}
H =\frac{1}{\alpha\big(Q_1^2+Q_2^2\big)+\beta Q_1+\gamma}\left(P_1^2+P_2^2+\frac{P_3^2}{16Q_2^2}\right).
\end{gather}
Since $\big\{K_2^2+K_3^2,K_1\big\}=0$, we use this to build an invariant of (\ref{h4-HQ})
\begin{gather*}
K_2^2+K_3^2 = 4 \left(T_1^2+\frac{(2\alpha Q_1+\beta)^2}{16Q_2^2}P_3^2\right),\qquad T_1=2\alpha Q_2P_1-(2\alpha Q_1+\beta)P_2,
\end{gather*}
is a first order integral of the kinetic energy.

The change of coordinates
\begin{gather*}
\bar u=\alpha\big(Q_1^2+Q_2^2\big)+\beta Q_1,\qquad \bar v=\frac1{2\alpha}\arctan{\frac{2\alpha Q_1+\beta}{2\alpha Q_2}},\qquad \bar w=Q_3,
\end{gather*}
gives
\begin{gather*}
 T_1=p_{\bar v},\qquad
H=\frac1{\bar u+\gamma}\left(\big(4\alpha \bar u+\beta^2\big)p_{\bar u}^2+\frac{1}{4\alpha \bar u+\beta^2}p_{\bar v}^2+\frac{\alpha^2p_{\bar w}^2}{4\big(4\alpha \bar u+\beta^2\big)\cos^2{(2\alpha \bar v)}}\right).
\end{gather*}
With the final step in the transformation,
\begin{gather*}
u =\ln\left(\bar u+\frac{\beta^2}{4\alpha}\right) ,\qquad v =4\alpha \bar v, \qquad w =\bar w,
\end{gather*}
we obtain
\begin{gather*} 
 T_1=4\alpha p_v,\qquad H=\frac{16\alpha^2\mathrm{e}^{-u}}{4\alpha\mathrm{e}^{u}+4\alpha\gamma-\beta^2}
 \left(p_u^2+p_v^2+\frac{p_w^2}{64\cos^2{\frac v2}}\right),
\end{gather*}
which has the $D_3$ kinetic energy, with potential (equivalent to the ``$b_1$'' part of Case~B in~\cite{03-11}).

\subsubsection{The conformal algebra and quadratic integrals}

We can see that $K_1=h_4$ commutes with the 6 functions
\begin{gather*}
 K_2^2+K_3^2,\ e_2^2+e_3^2,\ f_2^2+f_3^2,\ e_1,\ h_1,\ \mbox{and}\ f_1,
\end{gather*}
which correspond to the $6$-dimensional conformal algebra of our 2D kinetic energy $H_0$. In the present coordinates, these take the form
\begin{gather*}
T_1 = p_v, \qquad T_2 = {\rm e}^{-\frac{1}{2}u} \left(p_u \sin \frac{v}{2} +p_v \cos \frac{v}{2}\right), \qquad T_3 = {\rm e}^{-\frac{1}{2}u} \left(p_u \cos \frac{v}{2} -p_v \sin \frac{v}{2}\right), \\
T_4 = p_u,\qquad T_5 = {\rm e}^{\frac{1}{2}u} \left(p_u \sin \frac{v}{2} -p_v \cos \frac{v}{2}\right), \qquad T_6 = {\rm e}^{\frac{1}{2}u} \left(p_u \cos \frac{v}{2} +p_v \sin \frac{v}{2}\right).
\end{gather*}
The element $T_1$ is an {\em invariant} of the kinetic energy, but the full expression for $K_2^2+K_3^2$ gives a quadratic integral for~$H$:
\begin{gather*}
J_1 = T_1^2+\frac{1}{64} \operatorname{sech}^2 \frac{v}{2}\, p_w^2.
\end{gather*}
We also have that $\{F_1,K_1\}= 0$ (see Table~\ref{Tab:h4-Pij}), giving us
\begin{gather*}
J_2 = T_2^2 +\left(\beta^2-4 \alpha {\rm e}^u \sin^2 \frac{v}{2}\right) H.
\end{gather*}

\subsection[Reduction using the isometry $K_2=\alpha h_2+\beta e_2$]{Reduction using the isometry $\boldsymbol{K_2=\alpha h_2+\beta e_2}$}

To transform $K_2=\alpha h_2+\beta e_2$ to $P_3$ and to eliminate the $P_1 P_2$ term, we choose coordinates
\begin{gather*}
 Q_1=q_3,\qquad Q_2=\sqrt{\alpha q_1^2+\alpha q_2^2+\beta q_1},\qquad Q_3=-\frac1{2\alpha}\arctan\left(\frac{2\alpha q_1+\beta}{2\alpha q_2}\right),
\end{gather*}
giving
\begin{gather} \label{h4K2-HQ}
H = \frac{1}{\alpha Q_1^2+Q_2^2+\gamma}\left(P_1^2+\frac{4\alpha Q_2^2+\beta^2}{4Q_2^2}P_2^2+\frac{P_3^2}{4\alpha Q_2^2+\beta^2}\right).
\end{gather}
Since $\big\{K_2^2+K_3^2+\frac{\alpha^2}{4} K_1^2,K_2\big\}=0$, we use this to build an invariant of~(\ref{h4K2-HQ})
\begin{gather*}
K_2^2+K_3^2+\frac{\alpha^2}{4} K_1^2 = 4 \left(T_1^2+\frac{\big(4\alpha^2 Q_1^2+4\alpha Q_2^2+\beta^2\big)}{4\alpha Q_2^2+\beta^2}P_3^2\right),
\end{gather*}
where $T_1=\frac{\sqrt{4\alpha Q_2^2+\beta^2}}{Q_2}(Q_2P_1-\alpha Q_1P_2)$ is a first order integral of the kinetic energy.

The change of coordinates,
\begin{gather*}
 \bar u=\alpha Q_1^2+Q_2^2,\qquad \bar v=\frac{1}{2\alpha}\arctan{\frac{2\alpha Q_1}{\sqrt{4\alpha Q_2^2+\beta^2}}},\qquad \bar w=Q_3,
\end{gather*}
gives
\begin{gather*} 
T_1=p_{\bar v},\qquad H=\frac1{\bar u+\gamma}\left(\big(4\alpha \bar u+\beta^2\big)p_{\bar u}^2+\frac{1}{4\alpha \bar u+\beta^2}p_{\bar v}^2+\frac{p_{\bar w}^2}{\big(4\alpha \bar u+\beta^2\big)\cos^2{(2\alpha \bar v)}}\right).
\end{gather*}
With the final step in the transformation,
\begin{gather*}
u=\ln \left(\bar u+\frac{\beta^2}{4\alpha}\right) ,\qquad v=4\alpha \bar v,\qquad w= \bar w,
\end{gather*}
we obtain
\begin{gather*} 
 T_1=4\alpha p_v,\qquad H=\frac{16\alpha^2\mathrm{e}^{-u}}{4\alpha\mathrm{e}^{u}+4\alpha\gamma-\beta^2}
 \left(p_u^2+p_v^2+\frac{p_w^2}{16 \alpha^2\cos^2{\frac v2}}\right),
\end{gather*}
which has the $D_3$ kinetic energy, with {\it the same} potential (equivalent to the~``$b_1$'' part of Case~B in~\cite{03-11}).

\section{Conclusions}

The purpose of this paper was to extend the method developed in \cite{f18-4} from 2 to 3 degrees of freedom. The original motivation for~\cite{f18-4} was to understand Darboux--Koenigs super-integrable systems, which are characterised as having {\em exactly one} first order integral and 2 second order integrals. These are therefore associated with {\it non}-constant curvature metrics, which, in 2 dimensions, are necessarily conformally flat. The method of~\cite{f18-4} exploited the conformal flatness by building first integrals from conformal symmetries.

In the current paper we considered 3-dimensional conformally flat metrics and built integrals from the corresponding conformal algebra (\ref{g1234}). We initially assumed {\em one} first order integral $K$ and three quadratic ones $F_i$, giving us five integrals in all. A further requirement was involutivity of three of these and independence of them all, to give us a super-integrable system. These requirements are the obvious generalisation of the Darboux--Koenigs systems to 3 degrees of freedom.

The classification of such systems is a difficult task. As can be seen in the examples, we could start with a {\em different collection} of quadratic integrals to arrive at the {\em same} Hamiltonian (in particular, for the cases with larger symmetry algebras). However, the structure of the conformal algebra (\ref{g1234}), with its involutive automorphisms, enabled us to limit the starting choice of Killing vector to either $e_1$, $h_1$ or $h_4$, thus reducing the overall number of choices. We did not tackle the classification problem, but presented a number of interesting, multi-parameter examples, which, upon restriction, led to systems with a higher degree of symmetry, which were then able to be reduced to 2 degrees of freedom. Almost all our examples, upon reduction, had {\em one} linear and {\em two} quadratic first integrals, so were inevitably in the Darboux--Koenigs classification. However, the third dimension is ``remembered'' through the addition of a potential function, all of which were just special cases of those classified in \cite{03-11,02-6}. Furthermore, when there existed two inequivalent symmetries (with respect to involutions of the system under study) we were able to reduce in {\em two different} ways, leading to the introduction of two {\em different} potentials, and the two potentials could be related through a transformation in the 3-dimensional space. This means that the potentials in the classification of \cite{03-11,02-6} are not {\em fully} independent, {\it if} we allow transformations which take us out of the 2-dimensional space.

We see from the results of Section \ref{sec:F1=e2f2+red}, that {\em different parameter restrictions} of the 3-parameter Hamiltonian with conformal factor (\ref{phipsi}) can reduce to different Darboux--Koenigs cases. This means that the Hamiltonian (\ref{H3d-e1}), with $\varphi$ given by (\ref{phipsi}), simultaneously contains both Darboux--Koenigs cases $D_2$ and $D_4$. Furthermore, since there are only 4 metrics in the Darboux--Koenigs classification, we inevitably arrive at the {\em same} metric, from two or more {\em different} 3-dimensional metrics (see the examples of Sections~\ref{sec:F1=e2f2+red:gamma=0} and~\ref{F1=e2h2+:alpha=0-red}).

Whilst most of our examples (before reduction) possessed a 3-dimensional isometry algebra, the Hamiltonian~(\ref{F1=e2h2+:beta=0:H}) had only {\em one} first order integral, so the reduced metric had no Killing vectors at all. Nevertheless, we were able to show that the reduced Hamiltonian was maximally super-integrable, possessing quadratic and quartic integrals, thus falling out of the Darboux--Koenigs class of metric.

This paper represents an important first step in our investigations into the use of the conformal algebra to build higher order first integrals. Clearly we need a better understanding of the underlying structure in order to tackle the important open problem of the classification of such higher-dimensional superintegrable systems and their reductions.

Reducing from 3 to 2 dimensions, we acquired 1 parameter potentials. It is likely that if we start from a 4-dimensional system, with a large enough isometry algebra, we could produce a~2-parameter restriction of the potentials given in \cite{03-11,02-6}. Perhaps the {\em full} potential could be recovered by going to high enough dimensions.

\subsection*{Acknowledgements}

This work was supported by NSFC (Grant No. 11871396) and NSF of Shaanxi Province of China (Grant No. 2018JM1005). APF thanks Boris Kruglikov for useful discussions on the ``gap problem''. We thank the referees and an editor for their useful remarks.

\pdfbookmark[1]{References}{ref}
\LastPageEnding


\begin{thebibliography}{99}
\footnotesize\itemsep=0pt

\bibitem{09-10}
Ballesteros A., Enciso A., Herranz F.J., Ragnisco O., Superintegrability on
 {$N$}-dimensional curved spaces: central potentials, centrifugal terms and
 monopoles, \href{https://doi.org/10.1016/j.aop.2009.03.001}{\textit{Ann. Physics}} \textbf{324} (2009), 1219--1233,
 \href{https://arxiv.org/abs/0812.1882}{arXiv:0812.1882}.

\bibitem{11-4}
Ballesteros A., Enciso A., Herranz F.J., Ragnisco O., Riglioni D., Quantum
 mechanics on spaces of nonconstant curvature: the oscillator problem and
 superintegrability, \href{https://doi.org/10.1016/j.aop.2011.03.002}{\textit{Ann. Physics}} \textbf{326} (2011), 2053--2073,
 \href{https://arxiv.org/abs/1102.5494}{arXiv:1102.5494}.

\bibitem{84-4}
Dubrovin B.A., Fomenko A.T., Novikov S.P., Modern geometry -- methods and
 applications, Vols.~1--3, Springer-Verlag, New York, 1984.

\bibitem{17-3}
Escobar-Ruiz M.A., Miller Jr. W., Toward a classification of semidegenerate
 3{D} superintegrable systems, \href{https://doi.org/10.1088/1751-8121/aa5843}{\textit{J.~Phys.~A: Math. Theor.}} \textbf{50}
 (2017), 095203, 22~pages, \href{https://arxiv.org/abs/1611.02977}{arXiv:1611.02977}.

\bibitem{f18-2}
Fordy A.P., A {K}aluza--{K}lein reduction of super-integrable systems,
 \href{https://doi.org/10.1016/j.geomphys.2018.05.014}{\textit{J.~Geom. Phys.}} \textbf{131} (2018), 210--219, \href{https://arxiv.org/abs/1801.02981}{arXiv:1801.02981}.

\bibitem{f18-4}
Fordy A.P., First integrals from conformal symmetries: Darboux--Koenigs metrics
 and beyond, \href{https://arxiv.org/abs/1804.06904}{arXiv:1804.06904}.

\bibitem{f18-1}
Fordy A.P., Huang Q., Poisson algebras and 3{D} superintegrable {H}amiltonian
 systems, \href{https://doi.org/10.3842/SIGMA.2018.022}{\textit{SIGMA}} \textbf{14} (2018), 022, 37~pages,
 \href{https://arxiv.org/abs/1708.07024}{arXiv:1708.07024}.

\bibitem{74-7}
Gilmore R., Lie groups, {L}ie algebras, and some of their applications, Wiley,
 New York, 1974.

\bibitem{03-11}
Kalnins E.G., Kress J.M., Miller Jr. W., Winternitz P., Superintegrable systems
 in {D}arboux spaces, \href{https://doi.org/10.1063/1.1619580}{\textit{J.~Math. Phys.}} \textbf{44} (2003), 5811--5848,
 \href{https://arxiv.org/abs/math-ph/0307039}{arXiv:math-ph/0307039}.

\bibitem{02-6}
Kalnins E.G., Kress J.M., Winternitz P., Superintegrability in a
 two-dimensional space of nonconstant curvature, \href{https://doi.org/10.1063/1.1429322}{\textit{J.~Math. Phys.}}
 \textbf{43} (2002), 970--983, \href{https://arxiv.org/abs/math-ph/0108015}{arXiv:math-ph/0108015}.

\bibitem{72-5}
Koenigs G.X.P., Sur les g{\'e}od{\'e}siques a integrales quadratiques, in
 Le{\,c}ons sur la th{\'e}orie g{\'e}n{\'e}rale des surfaces, Vol.~4, Editor
 J.G.~Darboux, Chelsea Publishing, New York, 1972, 368--404.

\bibitem{14-7}
Kruglikov B., The D., The gap phenomenon in parabolic geometries,
 \href{https://doi.org/10.1515/crelle-2014-0072}{\textit{J.~Reine Angew. Math.}} \textbf{723} (2017), 153--215,
 \href{https://arxiv.org/abs/1303.1307}{arXiv:1303.1307}.

\bibitem{11-3}
Matveev V.S., Shevchishin V.V., Two-dimensional superintegrable metrics with
 one linear and one cubic integral, \href{https://doi.org/10.1016/j.geomphys.2011.02.012}{\textit{J. Geom. Phys.}} \textbf{61}
 (2011), 1353--1377, \href{https://arxiv.org/abs/1010.4699}{arXiv:1010.4699}.

\bibitem{13-2}
Miller Jr. W., Post S., Winternitz P., Classical and quantum superintegrability
 with applications, \href{https://doi.org/10.1088/1751-8113/46/42/423001}{\textit{J.~Phys.~A: Math. Theor.}} \textbf{46} (2013),
 423001, 97~pages, \href{https://arxiv.org/abs/1309.2694}{arXiv:1309.2694}.

\bibitem{17-4}
Valent G., Superintegrable models on {R}iemannian surfaces of revolution with
 integrals of any integer degree~({I}), \href{https://doi.org/10.1134/S1560354717040013}{\textit{Regul. Chaotic Dyn.}}
 \textbf{22} (2017), 319--352, \href{https://arxiv.org/abs/1703.10870}{arXiv:1703.10870}.

\end{thebibliography}
\end{document}